\documentclass[useAMS,usenatbib]{mn2e} 
\usepackage{url,graphics,times,array,amsmath,amssymb,ctable}

\newcommand{\CPDF}{P(<\tau_{\mathrm{eff}})}
\newcommand{\Mpc}{\mathrm{Mpc}}
\newcommand{\MFP}{\lambda_{\mathrm{mfp}}^{912}}

\newcommand{\Msun}{M_{\odot}}
\newcommand{\GammaHI}{\Gamma_{\mathrm{HI}}}

\newcommand{\dd}{\mathrm{d}}
\newcommand{\taueff}{\tau_{\mathrm{eff}}}

\newcommand{\HI}{H{\sc~i}}
\newcommand{\HII}{H{\sc~ii}}

\newcommand{\HeII}{He{\sc~ii}}
\newcommand{\HeIII}{He{\sc~iii}}

\title[Contribution of AGN to the ionizing background]{On the Contribution of Active Galactic Nuclei to the High-Redshift Metagalactic Ionizing Background}

\author[D'Aloisio et al.]{Anson D'Aloisio$^1$\thanks{Email:anson@u.washington.edu}, Phoebe R. Upton Sanderbeck$^1$, Matthew McQuinn$^1$, Hy Trac$^2$, \newauthor and Paul R. Shapiro$^3$ \\
$^1$Astronomy Department, University of Washington, Seattle, WA 98195 \\
$^2$McWilliams Center for Cosmology, Department of Physics, Carnegie Mellon University, Pittsburgh, PA 15213 \\
$^3$Astronomy Department and Texas Cosmology Center, University of Texas, Austin, TX 78712}

\begin{document}
\maketitle

\begin{abstract}
Motivated by the claimed detection of a large population of faint active galactic nuclei (AGN) at high redshift, recent studies have proposed models in which AGN contribute significantly to the $z>4$ \HI\ ionizing background. In some models, AGN are even the chief sources of reionization. If correct, these models would make necessary a complete revision to the standard view that galaxies dominated the high-redshift ionizing background.  It has been suggested that AGN-dominated models can better account for two recent observations that appear to be in conflict with the standard view: (1)  large opacity variations in the $z\sim 5.5$ \HI\ Ly$\alpha$ forest, and (2) slow evolution in the mean opacity of the \HeII\ Ly$\alpha$ forest.  Large spatial fluctuations in the ionizing background from the brightness and rarity of AGN may account for the former, while the earlier onset of \HeII\ reionization in these models may account for the latter.  Here we show that models in which AGN emissions source $\gtrsim 50\%$ of the ionizing background generally provide a better fit to the observed \HI\ Ly$\alpha$ forest opacity variations compared to standard galaxy-dominated models.  However, we argue that these AGN-dominated models are in tension with constraints on the thermal history of the intergalactic medium (IGM). Under standard assumptions about the spectra of AGN, we show that the earlier onset of \HeII\ reionization heats up the IGM well above recent temperature measurements. We further argue that the slower evolution of the mean opacity of the \HeII\ Ly$\alpha$ forest relative to simulations may reflect deficiencies in current simulations rather than favor AGN-dominated models as has been suggested.   
\end{abstract}


\section{Introduction}

Over the course of several decades, a standard view has emerged in which galaxies produced the dominant contribution of \HI\ ionizing photons at redshifts $z>4$ \citep[e.g.][]{1987ApJ...321L.107S,2009ApJ...703.1416F,2012ApJ...746..125H,2013MNRAS.436.1023B}. This view is based on numerous measurements of the AGN luminosity function which show a steep decline in the AGN abundance at $z>3$ \citep[e.g.][]{2006AJ....132..117F,2010AJ....139..906W,2013ApJ...768..105M,2015MNRAS.453.1946G}. Recently, the standard view has been challenged by some authors citing a large sample of faint AGN candidates reported by \citet[][G2015 hereafter]{2015A&amp;A...578A..83G}.  These objects were selected by searching for $X$-ray flux coincident with $z>4$ galaxy candidates in the CANDELS GOODS-South field -- a technique that in principle allows the detection of fainter AGN compared to prior selection methods.  If confirmed, this large population of faint AGN could make necessary a substantial revision of our understanding of the high-$z$ ionizing background, and possibly even of cosmological reionization.  Indeed, \citet{2015ApJ...813L...8M} showed that AGN with ionizing emissivities consistent with the G2015 measurements could reionize intergalactic \HI\ by $z\approx6$ without any contribution from galaxies (see also \citealt{2016MNRAS.457.4051K} and \citealt{2016arXiv160204407Y}). A distinguishing feature of their model is that \HeII\ reionization ends by $z \approx 4$, at least 500 million years earlier than in the standard scenario in which it ends at $z\approx 3$ \cite[see e.g.][and references therein]{2015arXiv151200086M}. 

An AGN-sourced ionizing background at $z>4$ potentially explains three puzzling observations of the IGM.  First, \HI\ Ly$\alpha$ forest\footnote{From here on, the term ``Ly$\alpha$ forest" refers to \HI.} measurements show that the \HI\ photoionization rate, $\GammaHI$, is remarkably flat over the redshift range $2<z<5$ \citep[see e.g.][]{2013MNRAS.436.1023B}.  This flatness is traditionally explained by invoking a steep increase in the escape fraction of galaxies with redshift, coinciding with the decline of the AGN abundance at $z>3$ \citep[e.g.][]{2012ApJ...746..125H}.  \citet{2015ApJ...813L...8M} showed that the slower evolution of the AGN emissivity claimed by G2015 can more naturally account for the observed flatness of $\GammaHI$ without appealing to a coincidental transition between the AGN and galaxy populations.  Second, measurements of the mean opacity of the \HeII\ Ly$\alpha$ forest at $z=3.1-3.3$ are lower than predictions from existing simulations of \HeII\ reionization, which use standard quasar emisisivity models with \HeII\ reionization ending at $z\approx3$ \citep{2014arXiv1405.7405W}.  \citet{2015ApJ...813L...8M} suggested that such low opacities are more consistent with an earlier onset of \HeII\ reionization driven by a large population of high-$z$ AGN.  Lastly, recent observations by \citet{2015MNRAS.447.3402B} show that the dispersion of opacities among coeval $50h^{-1}\Mpc$ segments of the Ly$\alpha$ forest increases rapidly above $z>5$, significantly exceeding the dispersion predicted by models that assume a uniform ionizing background.  In a companion paper \citep[][Paper I hereafter]{2016arXiv161102711D}, we show that accounting for this dispersion with spatial fluctuations in the ionizing background, under the standard assumption that galaxies are the dominant sources, requires that the mean free path of \HI\ ionizing photons be significantly shorter than observations and simulations indicate \citep[see also][]{2015MNRAS.447.3402B,2015arXiv150907131D}. Alternatively, models in which AGN source the ionizing background naturally lead to large fluctuations owing to the brightness and rarity of AGN. \citet{2015arXiv150501853C} showed with a ``proof-of-concept" model that rare sources with a space density of $\sim 10^{-6}~\Mpc^{-3}$, similar to the space density of $>L_*$ AGN in G2015, could generate large-scale ($\sim 50h^{-1}~\Mpc$) opacity variations substantial enough to account for the observed dispersion at $z=5.8$.  During the final preparation of this manuscript, \citet{2016arXiv160608231C} used more realistic models to show that a $\gtrsim 50\%$ contribution from AGN to the ionizing background is sufficient to account for the observed dispersion.  We reach a similar conclusion in this paper.  
 
The purpose of this paper is to further elucidate the implications of a large AGN population at high redshifts.  To this end we will discuss three observational probes of AGN-dominated models of the high-$z$ ionizing background: (1)  We will develop empirically motivated models of the Ly$\alpha$ forest in scenarios where AGN constitute a significant fraction of the background.  We will then use these models to assess the possible contribution of AGN to the $z>5$ Ly$\alpha$ forest opacity fluctuations; (2)   We will quantify the implications of these models for \HeII\ reionization and for the thermal history of the IGM -- a facet that has yet to be discussed in the literature; (3) We will discuss the interpretation of recent \HeII\ Ly$\alpha$ forest opacity measurements in the context of these models.   Foreshadowing, we will show that, while AGN-dominated models are indeed a viable explanation for the Ly$\alpha$ forest opacity measurements of \citet{2015MNRAS.447.3402B}, the models that best match the measurements are qualitatively inconsistent with constraints on the thermal history of the IGM under standard assumptions about the spectra of faint AGN.  We will further argue that the discrepancy between the opacities observed in the $z \approx3.1 - 3.3$ \HeII\ Ly$\alpha$ forest and those in current \HeII\ reionization simulations may reflect deficiencies in the simulations rather than favor AGN-dominated models.
    
The remainder of this paper is organized as follows.  In Section \ref{SEC:lumfunc} we present a comparison of the AGN luminosity function of G2015 to other measurements in the literature. Section \ref{SEC:opacityflucs} is dedicated to models of the Ly$\alpha$ forest opacity fluctuations, while \S \ref{SEC:thermalhistory} explores the impact of high-$z$ AGN on \HeII\ reionization and the thermal history of the IGM.  Section \ref{SEC:HeIIforest} discusses the interpretation of recent \HeII\ Ly$\alpha$ forest opacity measurements.  Finally, in \S \ref{SEC:conclusion} we offer closing remarks.  All distances are reported in comoving units unless otherwise noted.  We assume a flat $\Lambda$CDM cosmology with $\Omega_m=0.31$, $\Omega_b=0.048$, $h=0.68$, $\sigma_8=0.82$, $n_s=0.97$, and $Y_{\mathrm{He}}=0.25$, consistent with recent measurements \citep{2015arXiv150201589P}.

\section{The AGN Luminosity Function}
\label{SEC:lumfunc}

\begin{figure}
\begin{center}
\resizebox{8.5cm}{!}{\includegraphics{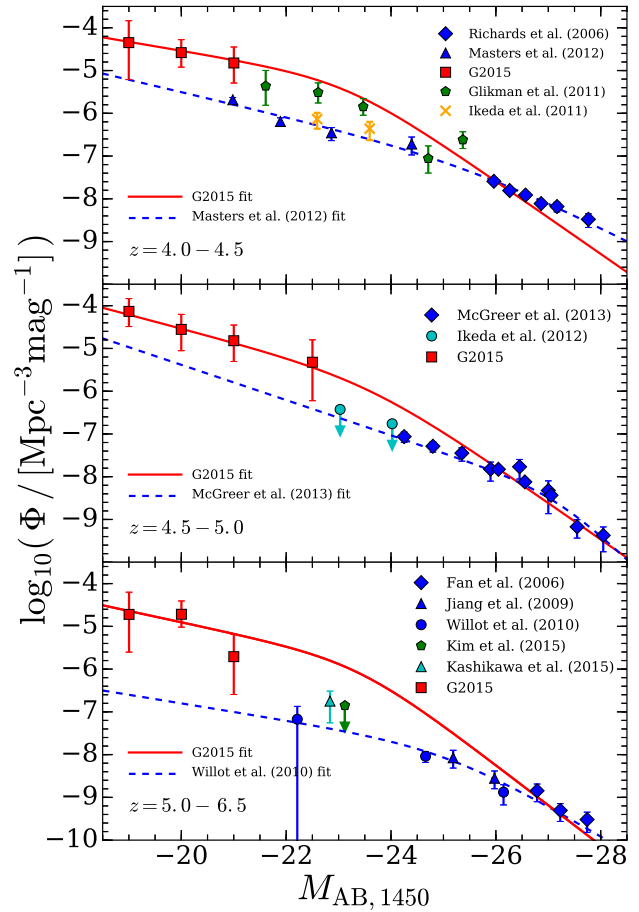}}
\end{center}
\caption{A comparison of the G2015 AGN luminosity function measurements to previously published measurements. The luminosity functions are expressed in terms of $M_{\rm AB,1450}$, the AB magnitude at a wavelength of 1450 \AA.   We adjust the redshifts of the previous measurements to the central redshifts of the G2015 bins by assuming that the luminosity function scales as $10^{-0.47z}$, a common approximation that is motivated by the observed evolution of the bright end of the luminosity function at $z=3-6$ \citep{2001AJ....121...54F}.  The G2015 luminosity function measurements are highly discrepant with previous measurements based on optically selected samples.} 
\label{FIG:lumfunc}
\end{figure}

\begin{figure}
\resizebox{8.5cm}{!}{\includegraphics{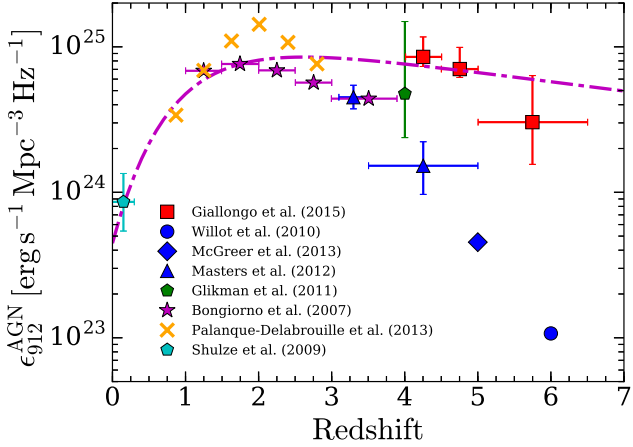}} 
\caption{Comoving emissivity of AGN at $h\nu = 1~\mathrm{Ry}~( \approx 13.6~\mathrm{eV})$.   For reference, the magenta/dot-dashed curve shows the model of \citet{2015ApJ...813L...8M}, which is motivated by the \citet{2015A&amp;A...578A..83G} emissivity measurements.  In this model, AGN emissions alone reionize \HI\ by $z\approx5.5$. See \citet{2015ApJ...813L...8M} for a similar comparison. }
\label{FIG:emissivity}
\end{figure}

We begin by comparing the AGN luminosity function measurements of G2015 to a compilation of other recent measurements based on optically selected samples (Fig. \ref{FIG:lumfunc}).  We adjust the redshifts of the other measurements to the central redshifts of the G2015 bins by assuming that the luminosity function scales as $10^{-0.47z}$, a common approximation that is motivated by the observed evolution of the bright end of the luminosity function over $z=3-6$ \citep{2001AJ....121...54F}.  Fig. \ref{FIG:lumfunc} shows that G2015's $X$-ray selected sample contains a much larger number of faint AGN than is expected from extrapolations of the past measurements.\footnote{We note that G2015's claims of a large number density of faint AGN ($M_{\mathrm{AB},1450} \approx -19$ to $-21$) may be supported by the AGN abundance found in the recent BlueTides simulation \citep{2016MNRAS.455.2778F}. However, a direct comparison is not possible because the simulation does not extend below $z=8$.}  To reconcile this discrepancy, G2015 argued that previous optical surveys missed a large fraction of high-$z$ AGN at intermediate luminosities ($M_{\mathrm{AB},1450} \approx -22$ to $-25$).  In fact, the G2015 fits shown in Fig. \ref{FIG:lumfunc} (red/solid lines) use only their faint AGN sample and a select subset of previous optical measurements at the bright end, which they argue are less prone to incompleteness.  Because of this, Fig. \ref{FIG:lumfunc} shows that extrapolating the G2015 luminosity function to intermediate luminosities yields a much larger AGN population than was found previously by the other surveys.  

G2015 assumed that the escape fractions of \HI\ ionizing photons for faint and intermediate luminosity AGN are similar to the measured values for quasars.  Under this assumption, their luminosity function translates to roughly a factor of ten greater contribution to the $z>4$ \HI\ ionizing background than previous measurements.  The data points in Fig. \ref{FIG:emissivity} translate the G2015 luminosity function, as well as others in the literature, to an ionizing emissivity of 1 Ry photons, 

\begin{equation}
\epsilon^{\rm AGN}_{912}= \int\dd L_\nu~\Phi(L_\nu)L_{912}(L_{\nu}), 
\label{EQ:em912}
\end{equation}
where $L_\nu$ is the specific luminosity, $\Phi(L_\nu)$ is the luminosity function, and $L_{912}(L_\nu)$ is the specific luminosity at 912 \AA\ (see \citealt{2015ApJ...813L...8M} for a similar comparison). We assume a power-law specific luminosity $\propto\nu^{-0.61}$ at wavelengths $>912$~\AA, consistent with the stacked quasar spectrum of \citet{2015MNRAS.449.4204L}. As in G2015, equation (\ref{EQ:em912}) assumes an escape fraction of unity, and we integrate integrate down to a specific luminosity of $L_* / 100$.  We note that much of the ionizing emissivity comes from the faint AGN population for the G2015 measurements.  For example, in their highest redshift bin, we find that $88\%, 71\%, 49\%$, and $30\%$ of the ionizing emissivity comes from AGN with $M_{\mathrm{AB},1450} > -24, -23, -22$, and $-21$, respectively.  If correct, the G2015 emissivity measurements could change the widely accepted view that galaxies were the chief sources of \HI\ reionization. The magenta dot-dashed curve shows the recently published model of \citet{2015ApJ...813L...8M}, in which AGN reionize \HI\ by $z\approx 5.5$ without {\it any} contribution from galaxies.  

We conclude this section by noting that \citet{2015MNRAS.448.3167W} searched for $z\gtrsim5$ AGN in the GOODS-South field -- the same field considered in G2015 -- using nearly the same data set and found no convincing candidates.   In the ensuing sections, we will adopt the point of view that the contribution of AGN to the $z>5$ ionizing background remains highly uncertain.  For the sake of exploring implications for the Ly$\alpha$ forest, we will consider several scenarios in which AGN contributed significantly to $z>5$ ionizing background.  

\begin{figure*}
\begin{center}
\resizebox{6.cm}{!}{\includegraphics{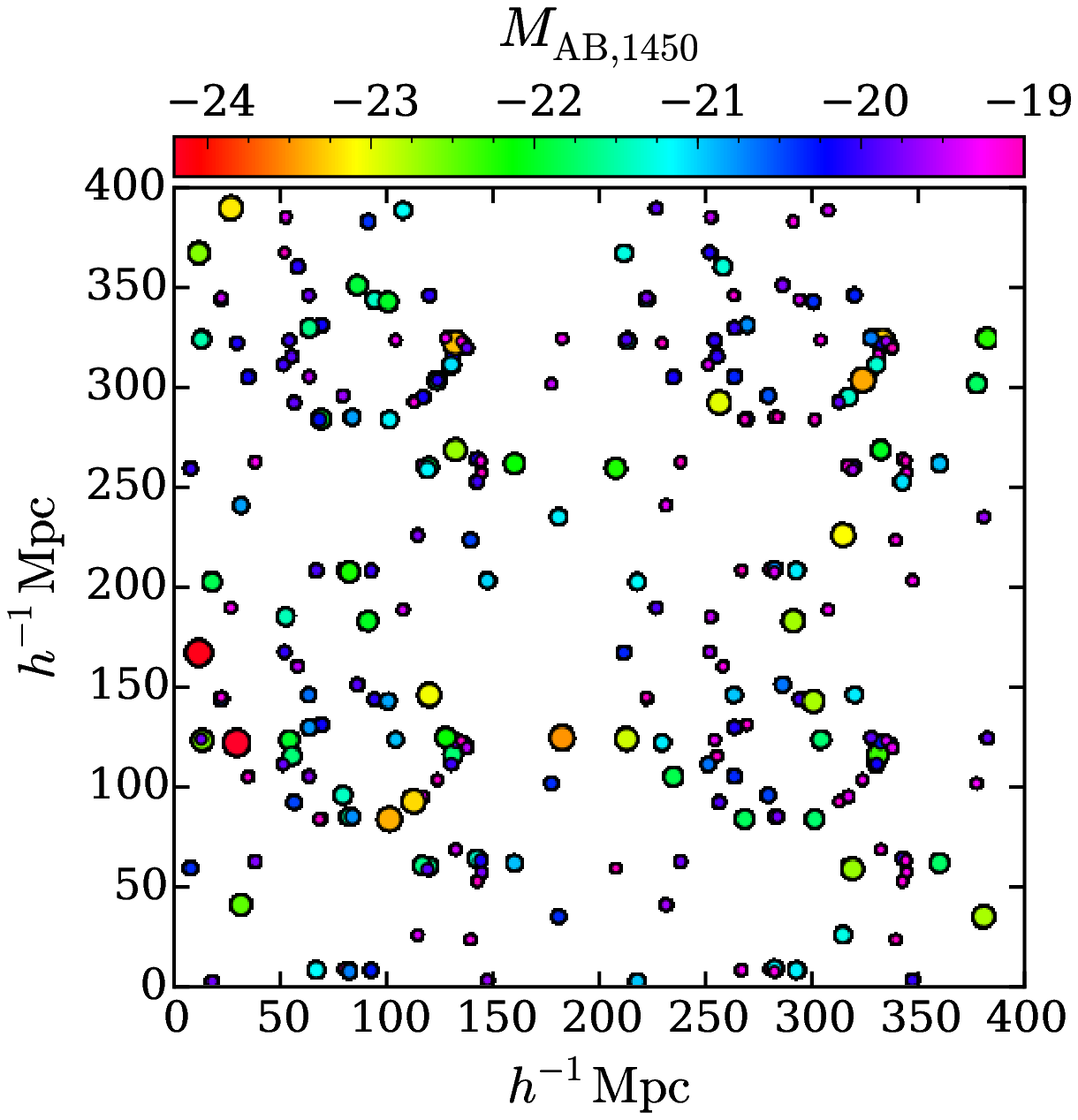}}
\hspace{-0.28cm}
\resizebox{6.cm}{!}{\includegraphics{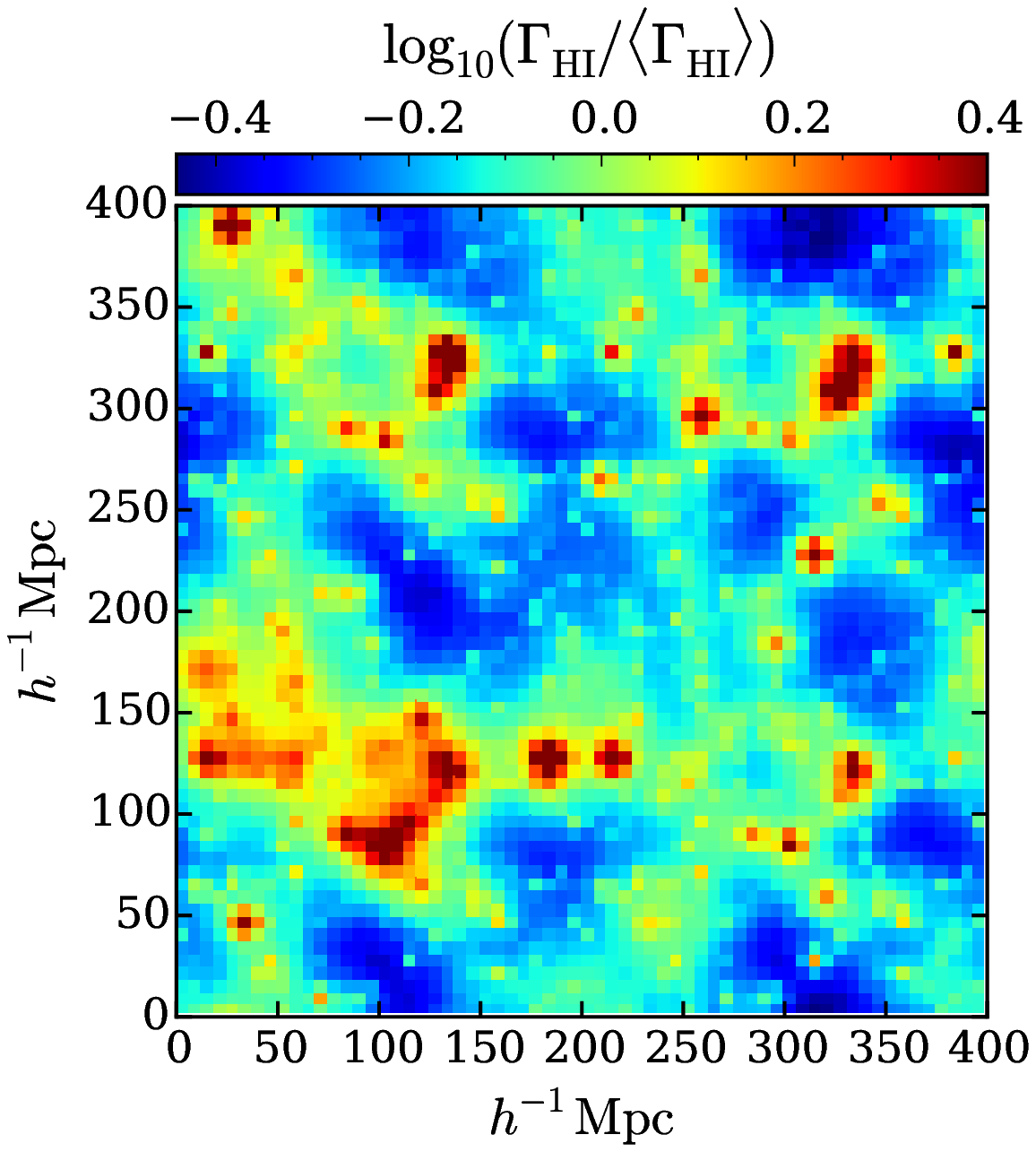}}
\hspace{-0.28cm}
\resizebox{6.cm}{!}{\includegraphics{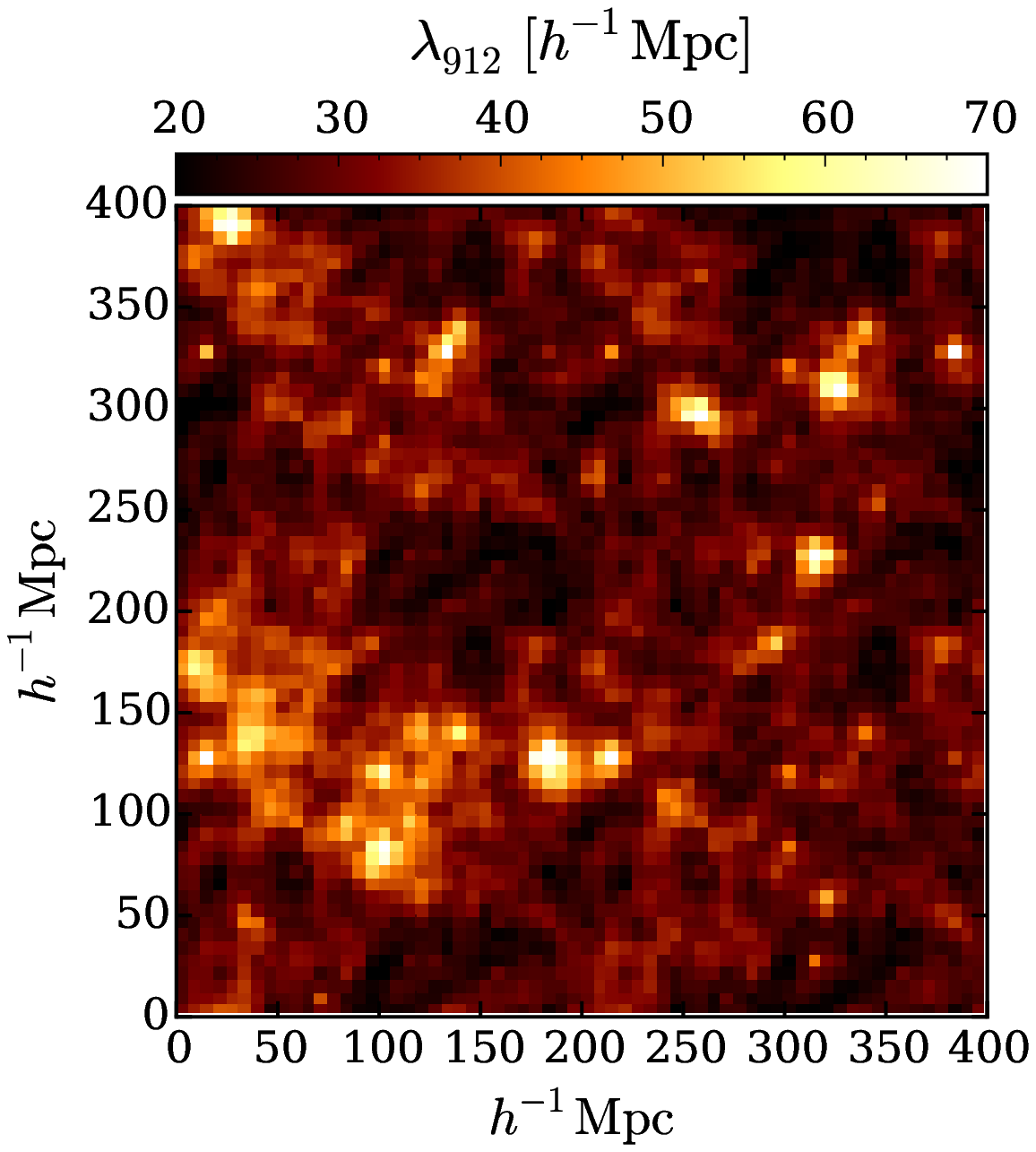}}
\end{center}
\caption{Fluctuations in the \HI\ ionizing background in one of our galaxies$+$AGN source models.  Here we show results from a model in which AGN emissions produce $50\%$ of the global average \HI\ photoionization rate. Left: a slice of thickness $20h^{-1}~\Mpc$ through the AGN source field.  We draw AGN from the luminosity function of \citet{2015A&amp;A...578A..83G} and randomly distribute them among the most massive galactic halos in our hydro simulation. To improve the statistics of our simulations of the ionizing background, we have tiled together eight of our hydro simulation boxes to achieve an effective $L_{\rm box}=400 h^{-1}~\Mpc$ before drawing AGN.  Middle: the \HI\ photoionization rate in a slice of thickness $6.25h^{-1}~\Mpc$, situated in the middle of the slice in the left panel. Right: the mean free path of 1 Ry photons in the same slice as in the middle panel.  The clustering of the AGN leads to large-scale variations in the mean free path that amplify fluctuations in the ionizing background.   }
\label{FIG:visB}
\end{figure*}

\section{Opacity Fluctuations in the High-Redshift Ly$\alpha$ Forest}
\label{SEC:opacityflucs}

The Ly$\alpha$ forest is the foremost observational probe of the metagalactic ionizing background and of the thermal state of the IGM.  At a given location of the forest, the Ly$\alpha$ optical depth, $\tau_{\rm Ly \alpha} \equiv -\ln F$ (where $F$ is the transmitted fraction of the quasar's flux) scales as $\tau_{\rm Ly \alpha} \propto T^{-0.7} \Delta_b/\GammaHI$.  Here, $T$ is the gas temperature, $\Delta_b$ is the gas density in units of the mean, and $\GammaHI$ is the photoionization rate which, in turn, scales in proportion to the strength of the local ionizing background.  Recently, \citet{2015MNRAS.447.3402B} showed that the observed transmission averaged over $50 h^{-1}~\Mpc$ segments of the forest displays an increasingly large scatter from sightline to slightline at redshifts approaching $z=6$.  At $z\gtrsim 5.5$ this scatter is too large to be explained by variations in the IGM density alone, indicating the presence of large spatial fluctuations in the ionizing background and/or in the IGM temperature \citep{2015arXiv150501853C, 2015ApJ...813L..38D, 2015arXiv150907131D, 2016arXiv160503183G}. 

\citet{2015ApJ...813L..38D} proposed a model for the latter in which the $z\sim 5.5$ opacity variations are generated largely by relic temperature fluctuations from patchy reionization.  In this scenario, the observed opacity fluctuation amplitude favors an extended but late-ending reionization process that was approximately half-complete by $z\approx 9$ and that ended at $z\approx 6$.  \citet{2015arXiv150907131D} argued that the opacity variations might not be tied to the reionization process at all; they might instead be a consequence of large fluctuations in the ionizing background well after the end of reionization.  In their model, these fluctuations are driven by the clustering of galaxies and by spatial variations in mean free path of ionizing photons.  The latter owes to the enhancement (suppression) of optically thick absorbers in voids (overdensities), where the local ionizing background is weaker (stronger).     

 { In paper I, we considered the model of \citet{2015arXiv150907131D} in further detail.  Under the standard assumption that galaxies dominate the background, we showed that accounting for the $z\approx5.5$ opacity fluctuations requires a short spatially averaged mean free path of $\langle \MFP \rangle \lesssim 15 h^{-1}~\Mpc$, a factor of $\approx$ 2 lower than is expected from extrapolations of observations at $z \leq 5.2$ \citep{2014MNRAS.445.1745W}.  We further showed that connecting this value with the measured mean free path at $z=5.2$ ($44 \pm 7 h^{-1}~\Mpc$, \citealt{2014MNRAS.445.1745W}) requires an unnatural factor of $\approx2$ decrease in the emissivity of the galaxy population in the $\approx100$ million years between $z=5.6$ and $z=5.2$.  Such a rapid evolution in the galaxy population would be surprising because the Hubble time scale is of order one billion years at these redshifts.  This is (to within a factor of a few) the time scale over which we should expect such a large change in the galaxy emissivity.\footnote{Recently, \citet{2016arXiv160503183G} claimed that the observed opacity fluctuations are well reproduced by ionizing background fluctuations in fully-coupled radiative transfer$+$hydrodynamics simulations with a standard galactic source model, and without the need for a small $\langle \MFP \rangle$.  However, in Paper I, some of us argue that their methodology is likely to overestimate the amplitude of $\taueff$ fluctuations.  }  However, we also identified a plausible solution to this problem.  We showed that, at $z\gtrsim 5$, the enhanced ionizing flux in the proximity zones of quasars can bias direct measurements of the mean free path higher than $\langle \MFP \rangle$ by up to a factor of two.  Such a large bias would reconcile the short values of $\langle \MFP \rangle$ that are required in the model of \citet{2015arXiv150907131D}.   } 

{ Given the large uncertainty in the magnitude of this bias (see discussion in Paper I), we explore an alternative model here. } One way to obviate the need for a short mean free path is to make the sources of the ionizing background in the post-reionization IGM much rarer and brighter than the faint, sub-$L_*$ galaxies that dominate in standard models. As \citet{2015arXiv150501853C} pointed out, the AGN of G2015, if they exist, are natural candidates for these rare sources.   In this section we shall consider an AGN-sourced ionizing background as a potential solution the problem of large opacity variations in the $z \approx 5.5$ Ly$\alpha$ forest.  In what follows, we focus on the effect of background fluctuations from AGN under the implicit assumption that residual temperature fluctuations from the \HI\ reionization process are weak at $z\lesssim 5.5$.  This assumption is appropriate for scenarios in which reionization ends significantly earlier than $z\approx 6$ and/or occurs rapidly.

\subsection{Numerical Methodology}
\label{SEC:methodology}   

Much of the numerical methodology for this work is described in paper I.  Here we briefly summarize the methodology and its extension to include AGN sources.  We used a modified version of the code of \citet{2004NewA....9..443T} to run a cosmological hydrodynamics simulation with box size $L_{\rm box} = 200 h^{-1}~\Mpc$, with $N_{\rm dm}= 2048^3$ dark matter particles and $N_{\rm gas}=2048^3$ gas cells.  This simulation does not include residual temperature fluctuations from \HI\ reionization (as noted above), nor does it include temperature fluctuations from \HeII\ reionization.  In Appendix \ref{SEC:model_uncertainties}, we discuss the impact of this on our conclusions.  To model the galaxy population, all dark mater halos with masses $M_{200}\geq 2\times 10^{10}h^{-1}~\Msun$ are populated with galaxies by abundance matching to the observed luminosity function of \citet{2015ApJ...803...34B}, using the scheme described in \citet{2015ApJ...813...54T}.  This minimum mass, which was chosen for completeness of the halo mass function, translates to a lower magnitude limit of $M_{\rm AB,1600} \approx -17.5$ at $z=5.5$, above the detection threshold for current observations.  We assume a constant escape fraction, which we tune in each model to match the observed mean transmission in the Ly$\alpha$ forest (see below for further details).

Since our models in this paper also include AGN as sources, which are much rarer than typical galaxies, we tile together eight of our hydro simulation boxes for an effective box size of $L_{\rm box}=400 h^{-1}~\Mpc$ before laying down AGN sources.  We randomly populate the most massive halos\footnote{For reference, the single most massive halo in our box at $z=5.2, 5.4$ and $5.6$ has $M_{200} = 8, 6,$ and $5.5\times10^{12}~h^{-1}\Msun$, respectively.} in this composite box with luminosities drawn from a rescaled version of the G2015 luminosity function, where the normalization is chosen to yield the desired level of contribution from AGN.  (See Appendix \ref{SEC:simLFs} for more details.)  We draw AGN down to a magnitude limit of $M_{\mathrm{AB},1450}=-19$, roughly the magnitude of the faintest candidates in G2015's sample.  We assume an AGN spectrum with $L_{\nu} \propto v^{-0.6}$ for $\lambda>912$ \AA, and $L_{\nu} \propto \nu^{-1.7}$ for $\lambda\leq912$ \AA, consistent with the quasar stack of \citet{2015MNRAS.449.4204L}.  We note that, while the escape fraction of \HI\ ionizing photons from quasars is likely close to unity, it is unclear whether this also holds for fainter AGN. One of our main goals is to assess the maximum possible contribution of AGN to the ionizing background, so we follow \citet{2015A&amp;A...578A..83G} and \citet{2015ApJ...813L...8M} in assuming a constant escape fraction of unity for all AGN.  

\begin{figure}
\begin{center}
\resizebox{8.8cm}{!}{\includegraphics{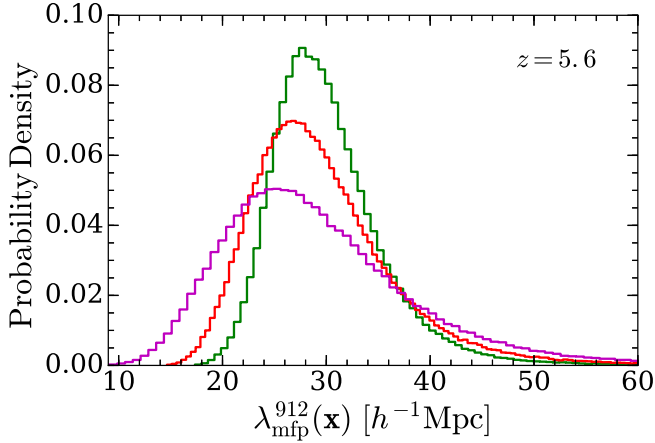}}
\end{center}
\caption{ { Distribution of mean free paths at $z=5.6$ in our $25\%$, $50\%$, and $90\%$ AGN models (narrowest to widest, respectively).  Fluctuations in the ionizing background are enhanced not only by a larger contribution from AGN, but also from the correspondingly larger fluctuations in the mean free path. }   }  
\label{FIG:MFPdist}
\end{figure}

After populating the sources, we post-process the hydro simulation with the model of \citet{2015arXiv150907131D} for the fluctuating ionizing background. This model self-consistently includes the effect of spatial variations in the mean free path. We compute $\GammaHI$ on a uniform grid with $N=64^3$, yielding a cell size of $\Delta x = 6.25 h^{-1}~\Mpc$.  In all models we set the spatially averaged mean free path to $\langle \MFP \rangle = 40, 35$, and $30~h^{-1}\Mpc$ at $z=5.2, 5.4$ and $5.6$, respectively, consistent with the extrapolation of the measurements of \citet{2014MNRAS.445.1745W}. In what follows, we will explore galaxies$+$AGN source models in which AGN constitute different fractions (i.e. $25, 50$ and $90 ~\%$) of the global average photoionization rate, $\langle \GammaHI \rangle$.  The left, middle, and right panels of Fig. \ref{FIG:visB} show slices through the AGN distribution, photoionization rate, and mean free path field at $z=5.6$ for our model where AGN contribute 50\% of $\langle \GammaHI \rangle$.  In the left panel we show all of the AGN in the slice, which has a thickness of $20h^{-1}~\Mpc$.  The right two panels have thicknesses of $6.25h^{-1}~\Mpc$, and are situated in the middle of the slice in the left panel. The fluctuations in $\GammaHI$ are enhanced by large-scale variations in the mean free path of ionizing photons, which is significantly longer in regions surrounding an overdensity of galaxies and/or several bright AGN. {  For reference, Fig. \ref{FIG:MFPdist} shows the distribution of $\MFP(\boldsymbol{x})$ at $z=5.6$ in our $25\%$, $50\%$, and $90\%$ AGN models (from narrowest to widest, respectively).  Fluctuations in $\GammaHI$ are enhanced not only by a larger contribution from AGN, but also from the correspondingly larger fluctuations in $\MFP(\boldsymbol{x})$. } 

Under the assumption of photoionization equilibrium (a good approximation in the post-reionization IGM), we combine these simulations of the fluctuating $\GammaHI$ field with our hydro simulations to produce synthetic Ly$\alpha$ forest spectra.  A typical way to quantify opacity fluctuations in the Ly$\alpha$ forest is to measure the distribution of effective optical depths, $\taueff \equiv -\ln \langle F \rangle_{L}$, for segments of the forest of length $L$, where $F$ is the continuum normalized flux.  The most recent measurements by \citet{2015MNRAS.447.3402B} adopted the convention of $L=50 h^{-1}~\Mpc$, so we will assume the same hereafter.  We construct the distribution of $\taueff$ from 4000 randomly oriented lines of sight. The spatially averaged photoionization rate, $\langle \GammaHI \rangle$, is a free parameter in our models that we fix by matching the observed transmission in the forest.  Specifically, we rescale $\GammaHI$ by a constant factor such that the mean value of $\langle F \rangle_{50}$ in our models is equal to the observed value in \citet{2015MNRAS.447.3402B}. For a given thermal state of the IGM, this normalization of $\langle \GammaHI \rangle$ also fixes the emissivities of the sources in our models.    

\subsection{Results}             

\begin{figure}
\begin{center}
\resizebox{8.8cm}{!}{\includegraphics{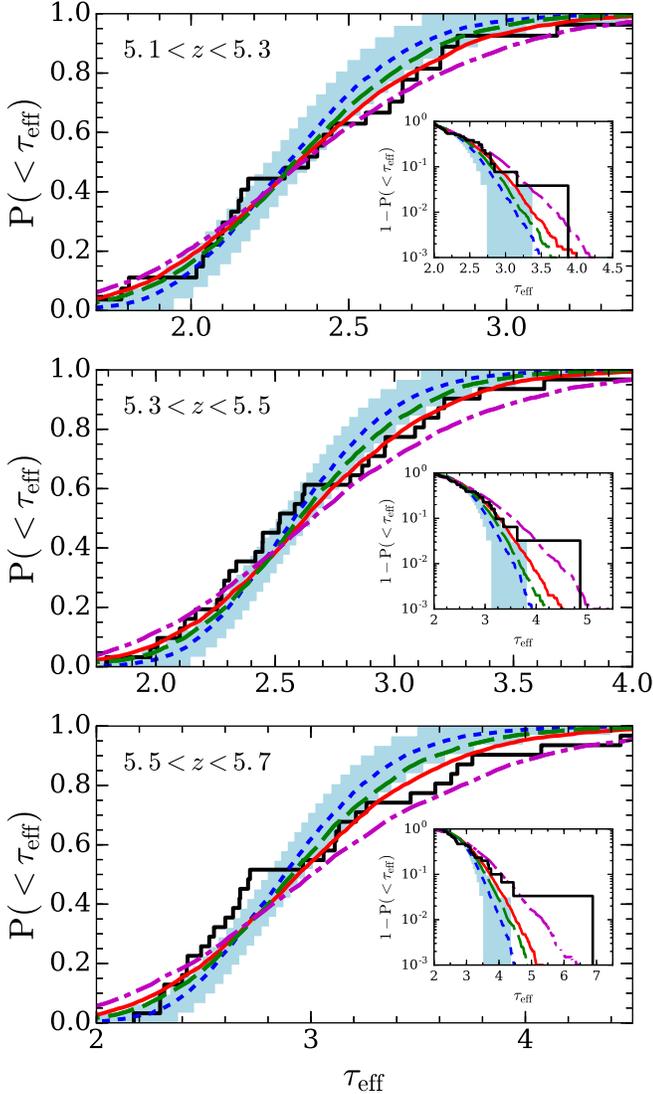}}
\end{center}
\caption{Opacity fluctuations in the $z>5$ Ly$\alpha$ forest in models where AGN contribute to the \HI\ ionizing background.  The green/long-dashed, red/solid, and magenta/dot-dashed curves correspond to models in which AGN emissions produce $25\%$, $50\%$ and $90\%$ of the global average \HI\ photoionization rate, respectively.  The black histograms show the cumulative probability distribution of $\taueff$ measured in $50h^{-1}\Mpc$ segments of the forest by \citet{2015MNRAS.447.3402B}. For reference, the blue/short-dashed curves correspond to a model in which only galaxies contribute to the ionizing background.  The light blue shading shows $90\%$ confidence regions estimated by bootstrap sampling of this model. For all models we assume our fiducial values for the spatially averaged mean free path,  $\langle \MFP \rangle = 40, 35$, and $30~h^{-1}\Mpc$ at $z=5.2, 5.4$, and $5.6$, consistent with the extrapolation of measurements by \citet{2014MNRAS.445.1745W}.}  
\label{FIG:taueff_AGN}
\end{figure}

\begin{figure}
\begin{center}
\resizebox{8.8cm}{!}{\includegraphics{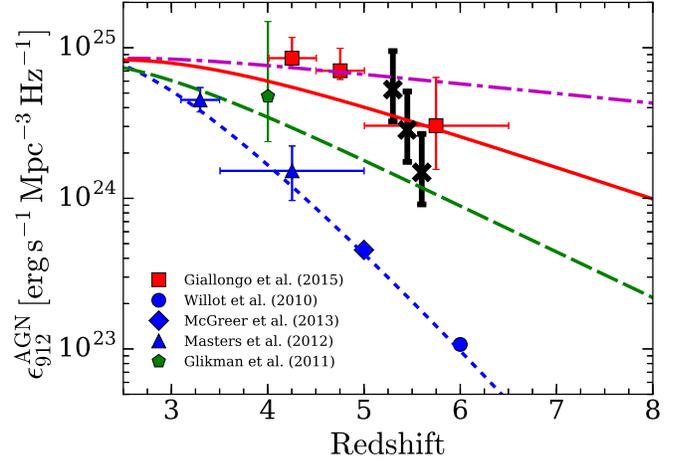}}
\end{center}
\caption{The \HI\ ionizing emissivities of AGN in our AGN-driven models (see Fig. \ref{FIG:taueff_AGN}).  From bottom to top, the black Xs show the $z=5.6$ ionizing emissivities in our $25\%$, $50\%$, and $90\%$ AGN models, respectively, while the corresponding error bars bracket the effects of the highly uncertain thermal state of the IGM.  For clarity, we have added a small horizontal offset between these data points.  The other data points show some of the high-$z$ measurements from Fig. \ref{FIG:emissivity}.  The curves correspond to parametric models that we use in \S \ref{SEC:thermalhistory} to explore the implications of an AGN-dominated ionizing background for \HeII\ reionization and the thermal history of the IGM (see main text for details).      }
\label{FIG:AGNemissivityB}
\end{figure}

\label{SEC:AGNmodels}

Fig. \ref{FIG:taueff_AGN} shows the cumulative probability distribution of the effective optical depth, $\CPDF$, in three models where AGN emissions constitute a significant fraction of the ionizing background. The green/long-dashed, red/solid and magenta/dot-dashed curves respectively correspond to scenarios in which AGN contribute $25\%$, $50\%$, and $90\%$ of $\langle \GammaHI \rangle$.\footnote{We note that the ionizing background fluctuations in these models are greatly enhanced by large-scale spatial variations in the mean free path.  To illustrate this, we have also considered models in which the mean free path is assumed to be uniform in space, with $\MFP=30h^{-1}~\Mpc$ at $z=5.6$.  (Note, all studies prior to \citealt{2015arXiv150907131D} that implemented a fluctuating ionizing background did so under the assumption of a uniform mean free path.)  In this case, the $50\%$ AGN model yields a $\CPDF$ that is very similar to the green/long-dashed curve in Fig. \ref{FIG:taueff_AGN}, and the $90\%$ AGN model is nearly identical to the red/solid curve. Thus, in these models it is crucial to account for spatial variations in the mean free path. }      The black histograms show the observational measurements of \citet{2015MNRAS.447.3402B} while, for reference, the blue/short-dashed curves show our fiducial model from Paper I in which only galaxies source the ionizing background.\footnote{In paper I we show that the $\taueff$ distribution in this ``galaxies-only" model is nearly identical to that in a model with spatially uniform $\GammaHI$.  }  The blue shading corresponds to $90\%$ confidence regions estimated by bootstrap sampling of this galaxies-only model (where we fix the mean $\langle F \rangle_{50}$ of each sample
to the observed value).  The $\taueff$ distribution in the $25\%$ AGN model is only mildly wider than in the galaxies-only model, and is consistently narrower than the observed distribution.  The $50\%$ AGN model provides a better match for much of the observed distribution at all redshifts, but the inset in the bottom panel of Fig. \ref{FIG:taueff_AGN} shows that even this model struggles to account for the highest opacity measurement at $z=5.6$.  On the other hand, while the $90\%$ AGN model can better account for the highest opacities, the distributions in this model are otherwise consistently wider than the observed distributions.  

 { An attractive feature of AGN-dominated models is that they do not require a short $\langle \MFP \rangle$ to account for (most of) the observed $\taueff$ dispersion. (Indeed, recall that we have assumed a $\langle \MFP \rangle(z)$ consistent with extrapolating the $z\lesssim 5.2$ measurements of \citealt{2014MNRAS.445.1745W}.)  Thus, unlike the galaxies-only model considered in paper I, they do not require an unnaturally rapid evolution in the emissivity of the galaxy population to be consistent with the mean free path measurements of \citealt{2014MNRAS.445.1745W}. In fact, we find that the redshift evolution of the galaxy emissivity is quite flat. For reference, in our $25\%$ AGN model, the mean emissivity from galaxies is $\langle \epsilon^{\rm gal}_{912}\rangle = 4.7, 4.7$, and $5.0\times10^{24}~\mathrm{erg~s^{-1}}~\Mpc^{-3}~\mathrm{Hz^{-1}}$ at $z=5.2, 5.4$ and $5.6$, respectively - a decrease of only $\approx 6\%$ between $z=5.6$ and $z=5.2$.  In Paper I, we argued that the mean free path measurements of \citealt{2014MNRAS.445.1745W} may be biased significantly by the enhanced ionizing flux in quasar proximity zones.  We note that such a bias could reduce the contribution from AGN that is required to account for the observed width of $\CPDF$.  }  

Let us now quantify in more detail the AGN contribution to the ionizing emissivities in these models, so we can compare them to the measurements of G2015. The value of the emissivity (which is determined by the observed mean transmission in the forest) depends on the thermal state of the IGM, which is highly uncertain at $z \gtrsim 5$.  In the AGN-dominated models, these uncertainties are made larger by the fact that the AGN population may begin reionizing \HeII\ earlier than in the standard scenario.  For a simple estimate of the upper and lower limits of these uncertainties, we parameterize the thermal state of the IGM with a temperature-density relation of the form $T(\Delta) = T_0 \Delta^{\gamma-1}$, where $T_0$ is the temperature at the cosmic mean density, and $\gamma$ specifies the logarithmic slope of the relation \citep{1997MNRAS.292...27H}.  Here we bracket the range of plausible thermal states with $(T_0,\gamma)=(20,000~\mathrm{K},1.2)$, and $(T_0,\gamma)=(5,000~\mathrm{K},1.6)$.  The former would be expected during or shortly after a reionization event, while the latter represent approximately the asymptotic values expected well after afterwards \citep{1997MNRAS.292...27H,2015MNRAS.450.4081P,2015arXiv150507875M,2015arXiv151105992U}.  Using these values, we rescale the AGN emissivity according to the scaling relations derived in \citet{2013MNRAS.436.1023B}, 

\begin{equation}
\epsilon_{912}(T_0,\gamma) \propto T_0^{-0.575} e^{ 0.7 \gamma},
\label{EQ:GammaRescale}
\end{equation} 
to obtain lower and upper limits, respectively.  At $z=5.6$, the ionizing emissivities in our $25 \%$, $50 \%$, and $90 \%$ AGN models are $\epsilon^{\rm AGN}_{912} = 1.5^{+1.2}_{-0.6}, 2.8^{+2.3}_{-1.1},$ and $5.3^{+4.2}_{-2.0} \times 10^{24}~\mathrm{erg~s^{-1} Hz^{-1} Mpc^{-1}}$, respectively.\footnote{See Paper I for a discussion of how we correct our emissivities for the effects of finite simulation resolution.  Here we also apply a crude correction to account for the fact that $\GammaHI$ receives a contribution from ionizing radiation produced during recombinations to the ground state of hydrogen \citep{1996ApJ...461...20H}.  At temperatures of $T\sim10,000$ K, the fraction of recombinations that produce an \HI\ ionizing photon is $(\alpha_{\rm A} - \alpha_{\rm B})/\alpha_{\rm A} \approx 40\%$, where $\alpha_{\rm A}$ and $\alpha_{\rm B}$ are the case A and B recombination coefficients, respectively. However, just a fraction of these photons are added to the ionizing background, since some of the recombinations will occur in optically thick absorbers, whose abundance is set by the \HI\ column density distribution.  For standard assumptions about the column density distribution, this fraction is roughly one half.  Thus we reduce our emissivities by $20\%$.  Previous studies that considered this radiative transfer effect in more detail have obtained similar corrections ranging from $\approx 10 - 30\%$ at these redshifts \citep{2009ApJ...703.1416F}.}

{ Fig. \ref{FIG:AGNemissivityB} compares the emissivities in our AGN-driven models to recent measurements in the literature. (The curves will be described in the next section.)  From bottom to top, the black Xs show the $z=5.6$ ionizing emissivities in our $25\%$, $50\%$, and $90\%$ AGN models, respectively.  For clarity, we have added a small horizontal offset between these data points.   Note that the emissivities in these models are of order the emissivity reported by G2015.  Even the 25\% AGN model has an emissivity that is a factor of $5-14$ higher than the value obtained by interpolating in redshift between the measurements of \citet{2013ApJ...768..105M} and \citet{2010AJ....139..906W}.  In the next section, we will discuss the implications of such a large ionizing emissivity from AGN for \HeII\ reionization and for the thermal history of the IGM.   }

\section{Helium Reionization and the Thermal History of the IGM}
\label{SEC:thermalhistory}

Compared to models in which galaxies dominate the ionizing background, the large AGN populations that we considered in the last section would likely emit many more photons with energies in excess of $4$ Ry.  These emissions would begin the reionization of intergalactic \HeII\ earlier than in the standard scenario.  In this section we explore the signature of this earlier \HeII\ reionization in the thermal history of the IGM.  

Below we consider four models with \HI\ ionizing emissivities of AGN given by the curves in Fig. \ref{FIG:AGNemissivityB}.  The form of the blue/short-dashed curve is given by equation (\ref{EQ:AGNemStandard}) in Appendix \ref{SEC:emissivitymodels}.  This model is representative of the standard scenario in which galaxies dominate $\epsilon_{912}$ at $z\gtrsim4$.  In contrast, the magenta/dot-dashed curve shows the model of \citet{2015ApJ...813L...8M}, in which AGN provide all of the ionizing emissivity necessary to reionize \HI\ by $z\approx5.5$.  { This model is representative of the $z=5.6$ emissivity in the $90\%$ AGN model from the last section.  The red/solid and green/long-dashed models are representative of the emissivities in the $50\%$ and $25\%$ AGN models, respectively. }  The form of these models is given by equation (\ref{EQ:modMH2015}). 

\begin{figure}
\begin{center}
\resizebox{8.2cm}{!}{\includegraphics{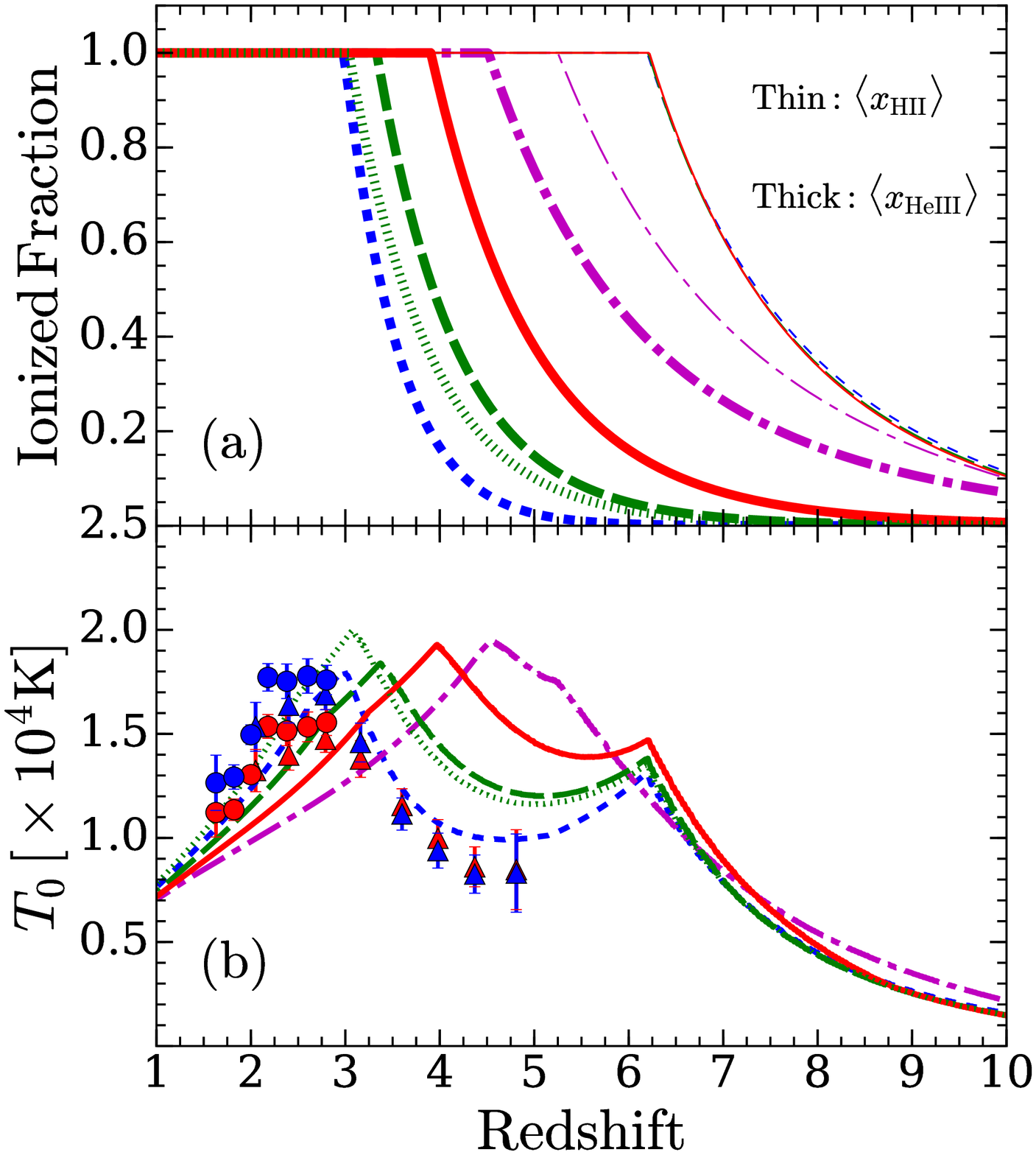}}
\resizebox{8.2cm}{!}{\includegraphics{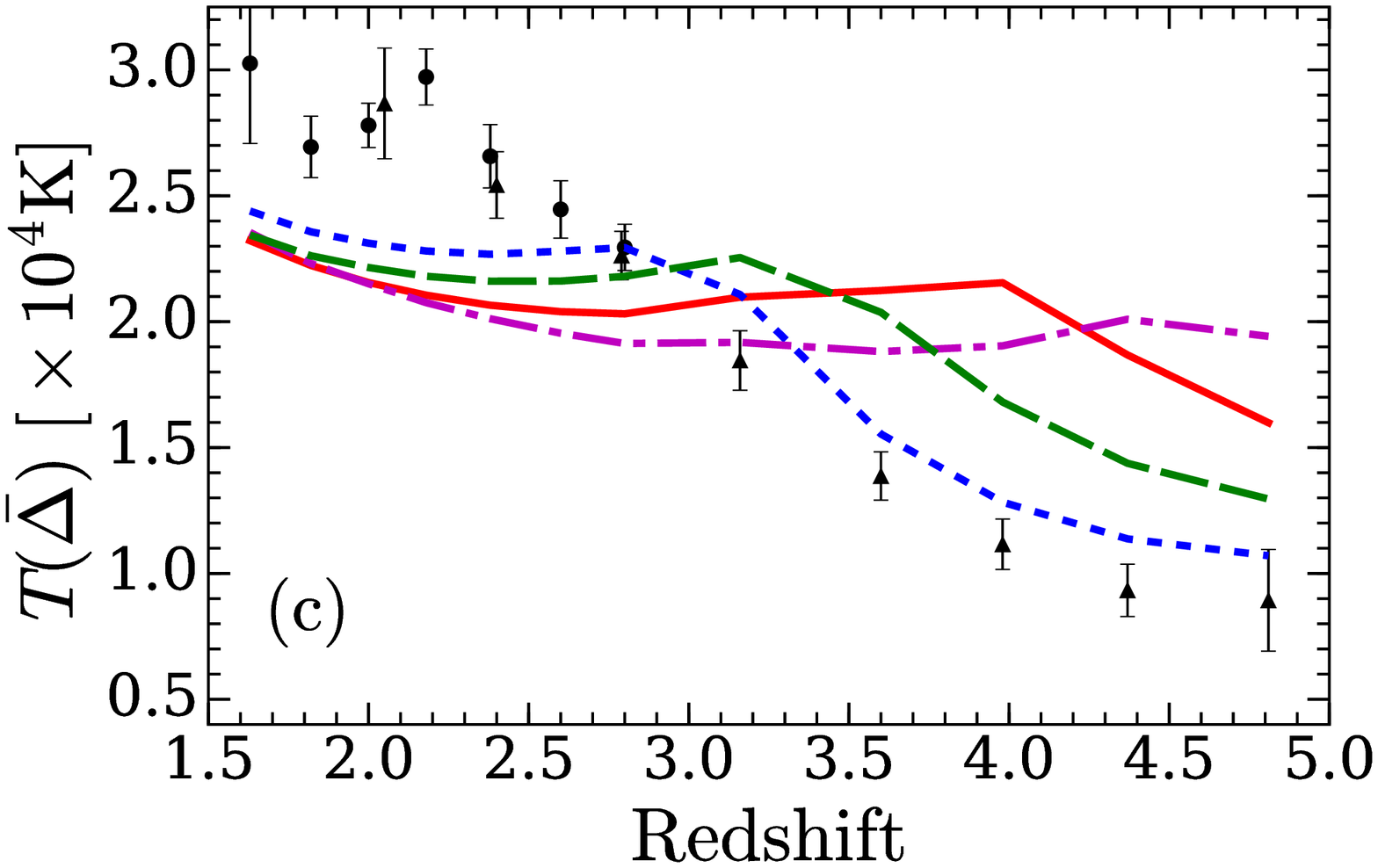}}
\end{center}
\caption{  The effect of early \HeII\ reionization on the thermal history of the IGM. Panel (a): global \HI\ and \HeII\ reionization histories (thin and thick, respectively) for models with \HI\ ionizing emissivities of AGN shown in Figure \ref{FIG:AGNemissivityB}. The cyan/dot-dashed curves correspond to the model of \citet{2015ApJ...813L...8M} in which AGN are assumed to be the only sources of \HI\ reionization.  For all other models, we tune the escape fraction of galaxies such that \HI\ reionization ends at $z\approx6$.  The dotted curve corresponds to a version of the green/long-dashed model in which \HeII\ reionization has been extended by increasing the recombination rate of \HeIII.  Panel (b): Temperature at the mean density of the universe. The data points show the Ly$\alpha$ forest temperature measurements of \citet{2011MNRAS.410.1096B} (triangles) and \citet{2014MNRAS.441.1916B} (circles).  The blue and red data points show these measurements extrapolated to the mean density using the temperature-density relation of the blue/short-dashed and red/solid models, respectively.  { For reference, panel (c) shows the comparison at the redshift-dependent ``optimal density," $\bar{\Delta}$, at which the temperature measurements were made. } }  
\label{FIG:temphistories}
\end{figure}

We solve the ionization balance equations to obtain the \HI\ and \HeII\ reionization histories, tuning the escape fraction of galaxies such that \HI\ reionization ends at $z\approx6$ in the three models that include galaxies.   For stellar sources we use the star formation rate density compiled by \citet{2015ApJ...802L..19R}.    The thin and thick sets of curves in  Fig. \ref{FIG:temphistories}a show $\langle x_{\mathrm{HII}}\rangle$ and $\langle x_{\mathrm{HeIII}}\rangle$, the volume-weighted mean ionized fractions of \HII\ and \HeIII\ in these models.  Our fiducial calculations assume constant clumping factors of $C_{\mathrm{HII}}=C_{\mathrm{HeIII}}=2$, approximately consistent with the values found in hydrodynamical simulations \citep[e.g.][]{2009MNRAS.394.1812P,mcquinn-Xray,2015ApJ...810..154K}, but we will discuss the effect of varying $C_{\mathrm{HeIII}}$ below. 

To model the thermal history of the IGM, we use a modified version of the two-zone approach of \citet{2015arXiv151105992U}.  In summary, we follow the thermal histories of an ensemble of gas parcels at different initial densities, using the Zel'dovich approximation for the parcels' dynamical histories.  Starting at $z^{\mathrm{HI}}_{\mathrm{start}}=12$, the redshift at which \HI\ reionization begins in our models, the parcels are photoheated by a uniform background of hard photons produced by the AGN population.  To delineate this hard background from the softer photons that are absorbed more locally at the boundary of ionization fronts, we include in the hard background only photons with mean free paths $>20~\Mpc$, but our results are insensitive to this choice.   We adopt a spectral index of $\alpha_{\rm QSO}=1.7$ for the hard background, and an index of $\alpha_{\rm bk}=1$ for the \HI\ ionizing background.  As \HI\ reionization progresses, the parcels are each instantaneously heated (in theory by a passing ionization front) to a temperature of $T=20,000$ K at a unique redshift, $z=z^{\mathrm{HI}}_{\mathrm{reion}}$, with cumulative probability distribution $\langle x_{\mathrm{HII}}\rangle$.  After $z^{\mathrm{HI}}_{\mathrm{reion}}$, the hard background continues heating by \HeII\ photoionizations until $z=z^{\mathrm{HeII}}_{\mathrm{reion}}$, at which times the parcels are instantaneously heated by $\approx8,000$ K [see eq. 12 of \citet{2015arXiv151105992U}]. This two zone model approximates the heating found in radiative transfer simulations of \HeII\ reionization \citep{2009ApJ...694..842M,2013MNRAS.435.3169C, 2016arXiv161002047L}, and is motivated by the steep dependence of the mean free path on photon energy.  

In models with a large AGN emissivity at high redshifts, $z^{\mathrm{HI}}_{\mathrm{reion}}$ and $z^{\mathrm{HeII}}_{\mathrm{reion}}$ are likely to be strongly correlated. To account for this correlation, we assume as an approximation the deterministic relation  

\begin{equation}
z^{\mathrm{HeII}}_{\mathrm{reion}}(z^{\mathrm{HI}}_{\mathrm{reion}})=z^{\mathrm{HI}}_{\mathrm{start}}\exp\left[-B_1(z^{\mathrm{HI}}_{\mathrm{start}}-z^{\mathrm{HI}}_{\mathrm{reion}})\right]-B_2.
\end{equation}  
Given the exact solution for $\langle x_{\mathrm{HII}}\rangle(z)$ in our models, we find that the values $B_1=(0.035,0.068,0.089,0.142) $ and $B_2=(6.8,4.5,3.2,0.05)$ yield excellent approximations to $\langle x_{\mathrm{HeIII}}\rangle(z)$ for the blue/short-dashed, green/long-dashed, red/solid, and magenta/dot-dashed models, respectively.  We find only minor differences in the resulting thermal histories if we assume the opposite limit in which $z^{\mathrm{HI}}_{\mathrm{reion}}$ and $z^{\mathrm{HeII}}_{\mathrm{reion}}$ are completely uncorrelated. 

The curves in Fig. \ref{FIG:temphistories}b show the IGM temperature at the mean density in our models. (The line styles match the emissivity histories in Fig. \ref{FIG:AGNemissivityB}.)  We compare them against the Ly$\alpha$ forest temperature measurements of \citet{2011MNRAS.410.1096B} (triangles) and \citet{2014MNRAS.441.1916B} (circles).  { It is important to note that these measurements are not at the mean density, but at a redshift-dependent optimal density, $\bar{\Delta}$ (see discussion in \citealt{2011MNRAS.410.1096B}).  The blue and red symbols show these measurements extrapolated to the mean density using the temperature-density relation of the blue/short-dashed and red/solid models, respectively.  For reference, Fig. \ref{FIG:temphistories}c shows the comparison at the original $\bar{\Delta}$ of the measurements.   These results show that the AGN-driven models considered in the last section are qualitatively inconsistent with the temperature measurements owing to the earlier reionization of \HeII\ and its associated photoheating.  Only the standard scenario (blue/short-dashed curve), in which \HeII\ reionization ends at $z\approx3$, provides a reasonable match to the high-redshift temperatures (see also \citealt{2015arXiv151105992U}). }

\begin{figure}
\begin{center}
\resizebox{8.4cm}{!}{\includegraphics{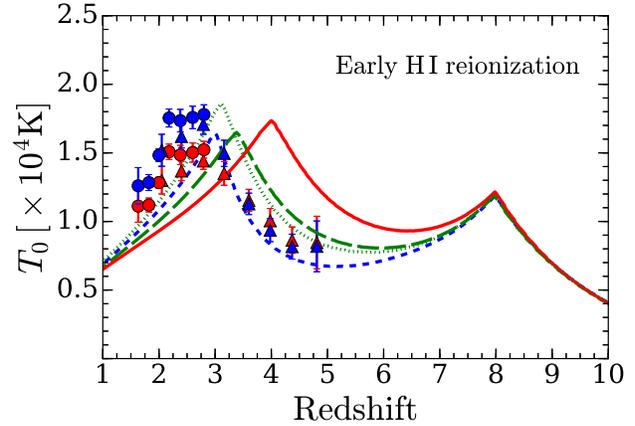}}
\end{center}
\caption{ { Same as in Fig. \ref{FIG:temphistories}b, except we assume that \HI\ reionization ends at $z\approx8$, much earlier than in our fiducial model (in which it ends at $z\approx6$).  Regardless of the details of \HI\ reionization, the earlier onset of \HeII\ reionization in AGN-dominated models renders them discrepant with the high-redshift temperature measurements. } } %As before, the triangles and circles show the measurements of \citet{2011MNRAS.410.1096B} and \citet{2014MNRAS.441.1916B}, respectively.   The blue and red symbols show these measurements extrapolated to the mean density using the temperature-density relation of the blue/short-dashed and red/solid models, respectively.    
\label{FIG:earlyHIreion}
\end{figure}

One way to reconcile the AGN-driven scenarios with observations is if \HeII\ reionization were significantly more extended than in our models.  To explore this possibility, we increased $C_{\mathrm{HeIII}}$ in the green/long-dashed model while holding \HI\ reionization fixed.  We found that delaying the end of \HeII\ reionization to $z\approx3$ in this model requires $C_{\mathrm{HeIII}}=4$, consistent with the highest values bracketed by the radiative transfer simulations of \citet{mcquinn-Xray}.  This model is depicted as the green/dotted curves in Fig. \ref{FIG:temphistories}. However, panel \ref{FIG:temphistories}b shows that the longer duration of \HeII\ reionization results in larger temperatures owing to the longer build-up of the hard radiation background from AGN and its associated photoheating.   Thus extending the duration of \HeII\ reionization appears to be in tension with the temperature measurements.  We note that this issue can be avoided if the fainter AGN, which account for most of the contribution to the \HI\ ionizing emissivity (see \S \ref{SEC:lumfunc}), have a much smaller escape fraction of \HeII\ ionizing photons.  However, a large escape fraction of \HI\ ionizing radiation would still be necessary to account for the observed Ly$\alpha$ forest opacity fluctuations considered in the last section. 

{ Finally, Fig. \ref{FIG:earlyHIreion} considers the impact of our assumptions about \HI\ reionization on our conclusions.  There, we adopt a scenario in which \HI\ reionization ends at $z\approx 8$, much earlier than in our fiducial model, such that the IGM starts out significantly colder at $z\approx6$.  (Note that we do not consider the AGN-only model of \citet{2015ApJ...813L...8M}, since its \HI\ reionization history is fixed by the AGN emissivity.)  We find that, regardless of the details of \HI\ reionization, the earlier onset of \HeII\ reionization still renders AGN-dominated models ($\gtrsim 50\%$ contribution from AGN to the ionizing background) discrepant with the high-redshift temperature measurements. }         

We conclude that, under standard assumptions about the spectra of AGN, it is difficult to reconcile AGN-driven models of the high-$z$ ionizing background -- especially scenarios with AGN emissivities as high as those claimed by G2015 -- with current observational constraints on the thermal history of the IGM.  On the other hand, recent observations of the \HeII\ Ly$\alpha$ forest indicate a lower effective optical depth at $z=3.1-3.3$ than is found in existing simulations of \HeII\ reionization \citep{2014arXiv1405.7405W}.  Some authors have argued that this low opacity is evidence for a more extended or earlier \HeII\ reionization \citep{2014arXiv1405.7405W,2015ApJ...813L...8M}.  In the next section we will address these claims.

\section{The \HeII\ Ly$\alpha$ Forest}
\label{SEC:HeIIforest}

Observations of the \HeII\ Ly$\alpha$ forest have been used to probe the timing and duration of \HeII\ reionization.  The \HeII\ Ly$\alpha$ forest displays Gunn-Peterson troughs at $z\geq2.7$ coeval with cosmic voids observed in the \HI\ Ly$\alpha$ forest.  The presence of these \HeII\ Ly$\alpha$ absorption troughs imply a \HeII\ fraction $\gtrsim10\%$ at the mean density, indicating that \HeII\ reionization was likely still in progress at $z\geq2.7$ \citep{2009ApJ...704L..89M}.  We note that this is qualitatively consistent with the blue/short-dashed model in Fig \ref{FIG:temphistories}, representative of the standard scenario in which AGN are subdominant at $z>4$.  On the other hand, \citet{2015ApJ...813L...8M} argued that a more extended (or earlier) \HeII\ reionization may be more consistent with the observed evolution in the \HeII\ Ly$\alpha$ effective optical depth reported by \citet{2014arXiv1405.7405W}.  At the highest redshifts probed by the data ($z=3.1-3.3$), \citet{2014arXiv1405.7405W} found lower \HeII\ Ly$\alpha$ optical depths than existing simulations of \HeII\ reionization (which use standard quasar emissivity models and hence have \HeII\ reionization end at $z\sim3$). 

Existing \HeII\ reionization simulations are likely to under-predict the optical depths of these measurements.  Early on during \HeII\ reionization, the regions that show significant transmission are likely the proximity regions of quasars.  The $z\gtrsim 3$ \HeII\ Ly$\alpha$ forest appears consistent with this picture, having large spans of no transmission punctuated by rare transmission peaks.  In this limit in which all of the \HeII\ Ly$\alpha$ transmission comes from  proximity regions, the transmission in the forest is given by
\begin{equation}
{\cal T} \approx \int \mathrm{d} L \; \Phi(L) V_{\rm eff}(L),
\label{eqn:HeII_trans}
\end{equation}
where $V_{\rm eff}(L)$ is the effective volume encapsulated by the average transmission profile around an individual quasar of luminosity $L$.  Using the 1D radiative transfer code presented in \citet{2015arXiv151103659K}, we find $V_{\rm eff} \propto L^{1.5}$, regardless of whether the quasar turns on in a \HeII\ region or in a \HeIII\ region, the same scaling expected for the classical proximity effect assuming ionization equilibrium.  However, we find that the proportionality changes by a factor of ten if quasar lifetimes vary from $10-100\;$Myr because of equilibration effects \citep{2015arXiv151103659K} and by an order unity factor if the quasar goes off in a \HeII\ region rather than in a \HeIII\ region.  This super-linear scaling means that $>L_*$ quasars contribute almost all of the transmission.

This simple model for the \HeII\ Ly$\alpha$ transmission illustrates two potential problems with the \HeII\ reionization simulations.  First, the simulations are typically limited to one or two quasar models, and the amount of transmission should be very sensitive to the quasar model.  We find that the transmission varies by a factor of ten depending on the lifetime that is assumed.  Indeed, most of the simulations in \citet{2009ApJ...694..842M} took the brightest quasars to have the shortest lifetimes, following \citet{2006ApJS..163....1H}, a choice that reduces the transmission relative to the more standard lightbulb models.   Second, simulations tend to miss quasars that occur at lower number densities than one per simulation volume.  For the volumes of previous simulations and using our model for the transmission, this omission leads to an underestimate of the transmission.  To mock up this effect in our model, we set the integrand in equation~(\ref{eqn:HeII_trans}) to zero at $ L \Phi= V^{-1}$, where $V$ is the simulation volume.  At $z=3.3$, we find that this cutoff reduces the transmission, ${\cal T}$, by a factor of $1.8$ for the $130h^{-1}\;$Mpc simulation box used in \citet{2009ApJ...694..842M} and by a factor of $2.0$ for the $100 h^{-1}~$Mpc box used in \citet{2013MNRAS.435.3169C}.  These factors increase to $2.3$ and $2.7$ respectively at $z=3.6$.   Thus, because of their limited volumes, existing \HeII\ reionization simulations underestimate the observed transmission (and hence $\tau_{\rm eff}$) in the \HeII\ forest in the limit that the transmission owes to isolated proximity effects.  While this limit does not apply at the end of \HeII\ reionization, when most locations see multiple quasars, it is likely to apply at the highest redshifts probed by the data, redshifts where the discrepancy reported by \citet{2014arXiv1405.7405W} arises.  Without larger volume simulations spanning a range of quasar lifetime models, the discrepancy between simulations and the \citet{2014arXiv1405.7405W} measurements should not be taken as evidence for a more extended or earlier \HeII\ reionization epoch.  Our simple model further suggests that the transmission in the high-redshift \HeII\ forest is less sensitive to the \HeII\ fraction than to the properties of quasar lifetimes.

\section{Conclusion}
\label{SEC:conclusion}

Several recent studies have proposed models in which AGN contribute significantly to (or even dominate) the $z\gtrsim 5$ \HI\ ionizing background. Among the main motivations for these models is that they may reconcile recent observations showing large opacity fluctuations in the $z\approx5.5$ \HI\ Ly$\alpha$ forest \citep{2015MNRAS.447.3402B}, and a slow evolution in the mean opacity of the $z\approx 3.1-3.3$ \HeII\ Ly$\alpha$ forest.  Here we have considered several facets of AGN-driven models of the ionizing background: (1) We have quantified the amplitude of $z\approx 5.5$ \HI\ Ly$\alpha$ forest opacity fluctuations in models with varying levels of contribution from AGN; (2) We have investigated the implications of these models for cosmological \HeII\ reionization and for the thermal history of the IGM; (3) We have discussed the interpretation of recent $\approx 3.1-3.3$ \HeII\ Ly$\alpha$ forest opacity measurements in the context of these models.  

We found that a model in which $\approx 50\%$ of the \HI\ ionizing background is sourced by AGN generally provides a better fit to the observed Ly$\alpha$ forest opacity fluctuation distribution across $z\approx 5-5.7$ compared to a model in which only galaxies source the background.  However, even this $50\%$ AGN model struggles to account for the highest opacity measurements. We found that doing so requires that essentially all ($\gtrsim 90\%$) of the ionizing background be produced by AGN, in a similar vein to the model proposed recently by \citet{2015ApJ...813L...8M}. These results are generally consistent with the findings of \citet{2016arXiv160608231C}, which appeared during the final preparations of this manuscript.  { An important caveat to this work is that we have adopted a model for the mean free path that is consistent with the recent measurements of \citealt{2014MNRAS.445.1745W}. In Paper I, we argued that these measurements may be biased significantly by the enhanced ionizing flux in quasar proximity zones.  We note that such a bias could reduce the contribution from AGN that is required to account for the observed amplitude of Ly$\alpha$ opacity fluctuations.}   

Since AGN are expected to have harder spectra than galaxies at \HeII\ ionizing energies of $\geq 4$ Ry, a unique prediction of AGN-driven models is that cosmological \HeII\ reionization occurs earlier than in the standard scenario \citep{2015ApJ...813L...8M}. Thus the impact of this reionization event on the thermal state of the IGM provides an independent way to constrain these models.  We showed that, under standard assumptions about the spectra of AGN, the earlier \HeII\ reionization heats the IGM well above the most recent temperature measurements, even for models in which AGN contribute modestly ($\approx 25\%$) to the \HI\ ionizing background. We are thus led to conclude that AGN-dominated models are disfavored by the temperature measurements.  Our results strongly disfavor the scenario proposed by \citet{2015ApJ...813L...8M} in which AGN emissions alone are enough to reionize intergalactic \HI\ by $z\approx5.5$.  

Finally, some authors have argued that the slow evolution in the mean opacity of the $z\approx 3.1-3.3$ \HeII\ Ly$\alpha$ forest implies that \HeII\ reionization ended earlier or was more extended than is expected in standard quasar source models \citep{2014arXiv1405.7405W,2015ApJ...813L...8M}.    We argued that the \HeII\ reionization simulations that were used to establish this discrepancy should over-predict the opacity at high redshifts.  We also argued that the opacity found in these simulations should be most dependent on the assumed quasar lightcurve model.  Thus the measurements of \citet{2014arXiv1405.7405W} are not necessarily in conflict with standard quasar models (and a late \HeII\ reionization).  Therefore, they should not yet be interpreted as evidence in favor of the abundant high-$z$ AGN models considered here.  
 
Combined with direct searches to better characterize the high-$z$ faint AGN population, future measurements of the Ly$\alpha$ forest opacity and of the IGM temperature will more tightly constrain the contribution of AGN to the ionizing background, as well as their contribution to reionization.   

\section*{Acknowledgements}
The authors acknowledge support from NSF grants AST1312724 and AST-1312724, from NASA through the Space Telescope
Science Institute grant HST-AR-13903.00, and from the NSF XSEDE allocations  TG-AST140087 and TG-AST150004.  H.T.  also acknowledges support from NASA grant ATP-NNX14AB57G.  P.R.S. acknowledges the support of U.S. NSF grant AST-1009799, NASA grant NNX11AE09G, NASA/JPL grant RSA Nos. 1492788 and 1515294.  A.D. thanks Fred Davies, Steven Furlanetto, Adam Lidz, Martin Haehnelt, and George Becker for useful discussions/correspondence.   The authors also thank the anonymous referee for comments and suggestions that improved this paper.    

\nocite{2012ApJ...755..169M,2011ApJ...728L..26G,2007A&amp;A...472..443B,2013A&amp;A...551A..29P,2009A&amp;A...507..781S,2006AJ....131.2766R,2011ApJ...728L..25I,2012ApJ...755..169M,2009AJ....138..305J,2015ApJ...798...28K,2012ApJ...756..160I, 2015ApJ...813L..35K}

\bibliography{./master}

\begin{thebibliography}{58}
\expandafter\ifx\csname natexlab\endcsname\relax\def\natexlab#1{#1}\fi

\bibitem[{{Becker} \& {Bolton}(2013)}]{2013MNRAS.436.1023B}
{Becker} G.~D., {Bolton} J.~S., 2013, \mnras, 436, 1023

\bibitem[{{Becker} {et~al}\mbox{.}(2011){Becker}, {Bolton}, {Haehnelt}, \&
  {Sargent}}]{2011MNRAS.410.1096B}
{Becker} G.~D., {Bolton} J.~S., {Haehnelt} M.~G., {Sargent} W.~L.~W., 2011,
  \mnras, 410, 1096

\bibitem[{{Becker} {et~al}\mbox{.}(2015){Becker}, {Bolton}, {Madau}, {Pettini},
  {Ryan-Weber}, \& {Venemans}}]{2015MNRAS.447.3402B}
{Becker} G.~D., {Bolton} J.~S., {Madau} P., {Pettini} M., {Ryan-Weber} E.~V.,
  {Venemans} B.~P., 2015, \mnras, 447, 3402

\bibitem[{{Boera} {et~al}\mbox{.}(2014){Boera}, {Murphy}, {Becker}, \&
  {Bolton}}]{2014MNRAS.441.1916B}
{Boera} E., {Murphy} M.~T., {Becker} G.~D., {Bolton} J.~S., 2014, \mnras, 441,
  1916

\bibitem[{{Bongiorno} {et~al}\mbox{.}(2007){Bongiorno}, {Zamorani},
  {Gavignaud}, {Marano}, {Paltani}, {Mathez}, {M{\o}ller}, {Picat}, \& {et
  al.}}]{2007A&amp;A...472..443B}
{Bongiorno} A. {et~al.}, 2007, \aap, 472, 443

\bibitem[{{Bouwens} {et~al}\mbox{.}(2015){Bouwens}, {Illingworth}, {Oesch},
  {Trenti}, {Labb{\'e}}, {Bradley}, {Carollo}, {van Dokkum}, {Gonzalez},
  {Holwerda}, {Franx}, {Spitler}, {Smit}, \& {Magee}}]{2015ApJ...803...34B}
{Bouwens} R.~J. {et~al.}, 2015, \apj, 803, 34

\bibitem[{{Chardin} {et~al}\mbox{.}(2015){Chardin}, {Haehnelt}, {Aubert}, \&
  {Puchwein}}]{2015arXiv150501853C}
{Chardin} J., {Haehnelt} M.~G., {Aubert} D., {Puchwein} E., 2015, \mnras, 453,
  2943

\bibitem[{{Chardin}, {Puchwein} \& {Haehnelt}(2016){Chardin}, {Puchwein}, \&
  {Haehnelt}}]{2016arXiv160608231C}
{Chardin} J., {Puchwein} E., {Haehnelt} M.~G., 2016, ArXiv e-prints

\bibitem[{{Compostella}, {Cantalupo} \& {Porciani}(2013){Compostella},
  {Cantalupo}, \& {Porciani}}]{2013MNRAS.435.3169C}
{Compostella} M., {Cantalupo} S., {Porciani} C., 2013, \mnras, 435, 3169

\bibitem[{{D'Aloisio} {et~al}\mbox{.}(2018){D'Aloisio}, {McQuinn}, {Davies}, \&
  {Furlanetto}}]{2016arXiv161102711D}
{D'Aloisio} A., {McQuinn} M., {Davies} F.~B., {Furlanetto} S.~R., 2018, \mnras, 473, 560

\bibitem[{{D'Aloisio}, {McQuinn} \& {Trac}(2015){D'Aloisio}, {McQuinn}, \&
  {Trac}}]{2015ApJ...813L..38D}
{D'Aloisio} A., {McQuinn} M., {Trac} H., 2015, \apjl, 813, L38

\bibitem[{{Davies} \& {Furlanetto}(2016)}]{2015arXiv150907131D}
{Davies} F.~B., {Furlanetto} S.~R., 2016, \mnras

\bibitem[{{Fan} {et~al}\mbox{.}(2006){Fan}, {Strauss}, {Becker}, {White},
  {Gunn}, {Knapp}, {Richards}, {Schneider}, {Brinkmann}, \&
  {Fukugita}}]{2006AJ....132..117F}
{Fan} X. {et~al.}, 2006, \aj, 132, 117

\bibitem[{{Fan} {et~al}\mbox{.}(2001){Fan}, {Strauss}, {Schneider}, {Gunn},
  {Lupton}, {Becker}, {Davis}, {Newman}, \& {et al.}}]{2001AJ....121...54F}
{Fan} X. {et~al.}, 2001, \aj, 121, 54

\bibitem[{{Faucher-Gigu{\`e}re} {et~al}\mbox{.}(2009){Faucher-Gigu{\`e}re},
  {Lidz}, {Zaldarriaga}, \& {Hernquist}}]{2009ApJ...703.1416F}
{Faucher-Gigu{\`e}re} C.-A., {Lidz} A., {Zaldarriaga} M., {Hernquist} L., 2009,
  \apj, 703, 1416

\bibitem[{{Feng} {et~al}\mbox{.}(2016){Feng}, {Di-Matteo}, {Croft}, {Bird},
  {Battaglia}, \& {Wilkins}}]{2016MNRAS.455.2778F}
{Feng} Y., {Di-Matteo} T., {Croft} R.~A., {Bird} S., {Battaglia} N., {Wilkins}
  S., 2016, \mnras, 455, 2778

\bibitem[{{Georgakakis} {et~al}\mbox{.}(2015){Georgakakis}, {Aird}, {Buchner},
  {Salvato}, {Menzel}, {Brandt}, {McGreer}, {Dwelly}, {Mountrichas}, {Koki},
  {Georgantopoulos}, {Hsu}, {Merloni}, {Liu}, {Nandra}, \&
  {Ross}}]{2015MNRAS.453.1946G}
{Georgakakis} A. {et~al.}, 2015, \mnras, 453, 1946

\bibitem[{{Giallongo} {et~al}\mbox{.}(2015){Giallongo}, {Grazian}, {Fiore},
  {Fontana}, {Pentericci}, {Vanzella}, {Dickinson}, {Kocevski}, \& {et
  al.}}]{2015A&amp;A...578A..83G}
{Giallongo} E. {et~al.}, 2015, \aap, 578, A83

\bibitem[{{Glikman} {et~al}\mbox{.}(2011){Glikman}, {Djorgovski}, {Stern},
  {Dey}, {Jannuzi}, \& {Lee}}]{2011ApJ...728L..26G}
{Glikman} E., {Djorgovski} S.~G., {Stern} D., {Dey} A., {Jannuzi} B.~T., {Lee}
  K.-S., 2011, \apjl, 728, L26

\bibitem[{{Gnedin}, {Becker} \& {Fan}(2016){Gnedin}, {Becker}, \&
  {Fan}}]{2016arXiv160503183G}
{Gnedin} N.~Y., {Becker} G.~D., {Fan} X., 2016, ArXiv e-prints

\bibitem[{{Haardt} \& {Madau}(1996)}]{1996ApJ...461...20H}
{Haardt} F., {Madau} P., 1996, \apj, 461, 20

\bibitem[{{Haardt} \& {Madau}(2012)}]{2012ApJ...746..125H}
{Haardt} F., {Madau} P., 2012, \apj, 746, 125

\bibitem[{{Hopkins} {et~al}\mbox{.}(2006){Hopkins}, {Hernquist}, {Cox}, {Di
  Matteo}, {Robertson}, \& {Springel}}]{2006ApJS..163....1H}
{Hopkins} P.~F., {Hernquist} L., {Cox} T.~J., {Di Matteo} T., {Robertson} B.,
  {Springel} V., 2006, \apjs, 163, 1

\bibitem[{{Hui} \& {Gnedin}(1997)}]{1997MNRAS.292...27H}
{Hui} L., {Gnedin} N.~Y., 1997, \mnras, 292, 27

\bibitem[{{Ikeda} {et~al}\mbox{.}(2012){Ikeda}, {Nagao}, {Matsuoka},
  {Taniguchi}, {Shioya}, {Kajisawa}, {Enoki}, {Capak}, {Civano}, {Koekemoer},
  {Masters}, {Morokuma}, {Salvato}, {Schinnerer}, \&
  {Scoville}}]{2012ApJ...756..160I}
{Ikeda} H. {et~al.}, 2012, \apj, 756, 160

\bibitem[{{Ikeda} {et~al}\mbox{.}(2011){Ikeda}, {Nagao}, {Matsuoka},
  {Taniguchi}, {Shioya}, {Trump}, {Capak}, {Comastri}, {Enoki}, {Ideue},
  {Kakazu}, {Koekemoer}, {Morokuma}, {Murayama}, {Saito}, {Salvato},
  {Schinnerer}, {Scoville}, \& {Silverman}}]{2011ApJ...728L..25I}
{Ikeda} H. {et~al.}, 2011, \apjl, 728, L25

\bibitem[{{Jiang} {et~al}\mbox{.}(2009){Jiang}, {Fan}, {Bian}, {Annis}, {Chiu},
  {Jester}, {Lin}, {Lupton}, {Richards}, {Strauss}, {Malanushenko},
  {Malanushenko}, \& {Schneider}}]{2009AJ....138..305J}
{Jiang} L. {et~al.}, 2009, \aj, 138, 305

\bibitem[{{Kashikawa} {et~al}\mbox{.}(2015){Kashikawa}, {Ishizaki}, {Willott},
  {Onoue}, {Im}, {Furusawa}, {Toshikawa}, {Ishikawa}, {Niino}, {Shimasaku},
  {Ouchi}, \& {Hibon}}]{2015ApJ...798...28K}
{Kashikawa} N. {et~al.}, 2015, \apj, 798, 28

\bibitem[{{Kaurov} \& {Gnedin}(2015)}]{2015ApJ...810..154K}
{Kaurov} A.~A., {Gnedin} N.~Y., 2015, \apj, 810, 154

\bibitem[{{Khaire} {et~al}\mbox{.}(2016){Khaire}, {Srianand}, {Choudhury}, \&
  {Gaikwad}}]{2016MNRAS.457.4051K}
{Khaire} V., {Srianand} R., {Choudhury} T.~R., {Gaikwad} P., 2016, \mnras, 457,
  4051

\bibitem[{{Khrykin} {et~al}\mbox{.}(2015){Khrykin}, {Hennawi}, {McQuinn}, \&
  {Worseck}}]{2015arXiv151103659K}
{Khrykin} I.~S., {Hennawi} J.~F., {McQuinn} M., {Worseck} G., 2015, ArXiv
  e-prints

\bibitem[{{Kim} {et~al}\mbox{.}(2015){Kim}, {Im}, {Jeon}, {Kim}, {Choi},
  {Hong}, {Hyun}, {Jun}, {Karouzos}, {Kim}, {Kim}, {Kim}, {Kim}, {Lee}, {Pak},
  {Park}, {Taak}, \& {Yoon}}]{2015ApJ...813L..35K}
{Kim} Y. {et~al.}, 2015, \apjl, 813, L35

\bibitem[{{La Plante} {et~al}\mbox{.}(2016){La Plante}, {Trac}, {Croft}, \&
  {Cen}}]{2016arXiv161002047L}
{La Plante} P., {Trac} H., {Croft} R., {Cen} R., 2016, ArXiv e-prints

\bibitem[{{Lusso} {et~al}\mbox{.}(2015){Lusso}, {Worseck}, {Hennawi},
  {Prochaska}, {Vignali}, {Stern}, \& {O'Meara}}]{2015MNRAS.449.4204L}
{Lusso} E., {Worseck} G., {Hennawi} J.~F., {Prochaska} J.~X., {Vignali} C.,
  {Stern} J., {O'Meara} J.~M., 2015, \mnras, 449, 4204

\bibitem[{{Madau} \& {Haardt}(2015)}]{2015ApJ...813L...8M}
{Madau} P., {Haardt} F., 2015, \apjl, 813, L8

\bibitem[{{Masters} {et~al}\mbox{.}(2012){Masters}, {Capak}, {Salvato},
  {Civano}, {Mobasher}, {Siana}, {Hasinger}, {Impey}, {Nagao}, {Trump},
  {Ikeda}, {Elvis}, \& {Scoville}}]{2012ApJ...755..169M}
{Masters} D. {et~al.}, 2012, \apj, 755, 169

\bibitem[{{McGreer} {et~al}\mbox{.}(2013){McGreer}, {Jiang}, {Fan}, {Richards},
  {Strauss}, {Ross}, {White}, {Shen}, {Schneider}, {Myers}, {Brandt}, {DeGraf},
  {Glikman}, {Ge}, \& {Streblyanska}}]{2013ApJ...768..105M}
{McGreer} I.~D. {et~al.}, 2013, \apj, 768, 105

\bibitem[{{McQuinn}(2009)}]{2009ApJ...704L..89M}
{McQuinn} M., 2009, \apjl, 704, L89

\bibitem[{{McQuinn}(2012)}]{mcquinn-Xray}
{McQuinn} M., 2012, \mnras, 426, 1349

\bibitem[{{McQuinn}(2016)}]{2015arXiv151200086M}
{McQuinn} M., 2016, \araa, 54, 313

\bibitem[{{McQuinn} {et~al}\mbox{.}(2009){McQuinn}, {Lidz}, {Zaldarriaga},
  {Hernquist}, {Hopkins}, {Dutta}, \&
  {Faucher-Gigu{\`e}re}}]{2009ApJ...694..842M}
{McQuinn} M., {Lidz} A., {Zaldarriaga} M., {Hernquist} L., {Hopkins} P.~F.,
  {Dutta} S., {Faucher-Gigu{\`e}re} C.-A., 2009, \apj, 694, 842

\bibitem[{{McQuinn} \& {Upton Sanderbeck}(2016)}]{2015arXiv150507875M}
{McQuinn} M., {Upton Sanderbeck} P.~R., 2016, \mnras, 456, 47

\bibitem[{{Palanque-Delabrouille} {et~al}\mbox{.}(2013){Palanque-Delabrouille},
  {Magneville}, {Y{\`e}che}, {Eftekharzadeh}, {Myers}, {Petitjean},
  {P{\^a}ris}, {Aubourg}, \& {et al.}}]{2013A&amp;A...551A..29P}
{Palanque-Delabrouille} N. {et~al.}, 2013, \aap, 551, A29

\bibitem[{{Pawlik}, {Schaye} \& {van Scherpenzeel}(2009){Pawlik}, {Schaye}, \&
  {van Scherpenzeel}}]{2009MNRAS.394.1812P}
{Pawlik} A.~H., {Schaye} J., {van Scherpenzeel} E., 2009, \mnras, 394, 1812

\bibitem[{{Planck Collaboration} {et~al}\mbox{.}(2015){Planck Collaboration},
  {Ade}, {Aghanim}, {Arnaud}, {Ashdown}, {Aumont}, {Baccigalupi}, {Banday},
  {Barreiro}, {Bartlett}, \& et~al.}]{2015arXiv150201589P}
{Planck Collaboration} {et~al.}, 2015, ArXiv e-prints

\bibitem[{{Puchwein} {et~al}\mbox{.}(2015){Puchwein}, {Bolton}, {Haehnelt},
  {Madau}, {Becker}, \& {Haardt}}]{2015MNRAS.450.4081P}
{Puchwein} E., {Bolton} J.~S., {Haehnelt} M.~G., {Madau} P., {Becker} G.~D.,
  {Haardt} F., 2015, \mnras, 450, 4081

\bibitem[{{Richards} {et~al}\mbox{.}(2006){Richards}, {Strauss}, {Fan}, {Hall},
  {Jester}, {Schneider}, {Vanden Berk}, {Stoughton}, {Anderson}, {Brunner},
  {Gray}, {Gunn}, {Ivezi{\'c}}, {Kirkland}, {Knapp}, {Loveday}, {Meiksin},
  {Pope}, {Szalay}, {Thakar}, {Yanny}, {York}, {Barentine}, {Brewington},
  {Brinkmann}, {Fukugita}, {Harvanek}, {Kent}, {Kleinman}, {Krzesi{\'n}ski},
  {Long}, {Lupton}, {Nash}, {Neilsen}, {Nitta}, {Schlegel}, \&
  {Snedden}}]{2006AJ....131.2766R}
{Richards} G.~T. {et~al.}, 2006, \aj, 131, 2766

\bibitem[{{Robertson} {et~al}\mbox{.}(2015){Robertson}, {Ellis}, {Furlanetto},
  \& {Dunlop}}]{2015ApJ...802L..19R}
{Robertson} B.~E., {Ellis} R.~S., {Furlanetto} S.~R., {Dunlop} J.~S., 2015,
  \apjl, 802, L19

\bibitem[{{Schulze}, {Wisotzki} \& {Husemann}(2009){Schulze}, {Wisotzki}, \&
  {Husemann}}]{2009A&amp;A...507..781S}
{Schulze} A., {Wisotzki} L., {Husemann} B., 2009, \aap, 507, 781

\bibitem[{{Shapiro} \& {Giroux}(1987)}]{1987ApJ...321L.107S}
{Shapiro} P.~R., {Giroux} M.~L., 1987, \apjl, 321, L107

\bibitem[{{Trac}, {Cen} \& {Mansfield}(2015){Trac}, {Cen}, \&
  {Mansfield}}]{2015ApJ...813...54T}
{Trac} H., {Cen} R., {Mansfield} P., 2015, \apj, 813, 54

\bibitem[{{Trac} \& {Pen}(2004)}]{2004NewA....9..443T}
{Trac} H., {Pen} U.-L., 2004, \na, 9, 443

\bibitem[{{Upton Sanderbeck}, {D'Aloisio} \& {McQuinn}(2016){Upton Sanderbeck},
  {D'Aloisio}, \& {McQuinn}}]{2015arXiv151105992U}
{Upton Sanderbeck} P.~R., {D'Aloisio} A., {McQuinn} M.~J., 2016, \mnras, 460,
  1885

\bibitem[{{Weigel} {et~al}\mbox{.}(2015){Weigel}, {Schawinski}, {Treister},
  {Urry}, {Koss}, \& {Trakhtenbrot}}]{2015MNRAS.448.3167W}
{Weigel} A.~K., {Schawinski} K., {Treister} E., {Urry} C.~M., {Koss} M.,
  {Trakhtenbrot} B., 2015, \mnras, 448, 3167

\bibitem[{{Willott} {et~al}\mbox{.}(2010){Willott}, {Delorme}, {Reyl{\'e}},
  {Albert}, {Bergeron}, {Crampton}, {Delfosse}, {Forveille}, {Hutchings},
  {McLure}, {Omont}, \& {Schade}}]{2010AJ....139..906W}
{Willott} C.~J. {et~al.}, 2010, \aj, 139, 906

\bibitem[{{Worseck} {et~al}\mbox{.}(2014{\natexlab{a}}){Worseck}, {Prochaska},
  {Hennawi}, \& {McQuinn}}]{2014arXiv1405.7405W}
{Worseck} G., {Prochaska} J.~X., {Hennawi} J.~F., {McQuinn} M.,
  2014{\natexlab{a}}, ArXiv e-prints

\bibitem[{{Worseck} {et~al}\mbox{.}(2014{\natexlab{b}}){Worseck}, {Prochaska},
  {O'Meara}, {Becker}, {Ellison}, {Lopez}, {Meiksin}, {M{\'e}nard}, {Murphy},
  \& {Fumagalli}}]{2014MNRAS.445.1745W}
{Worseck} G. {et~al.}, 2014{\natexlab{b}}, \mnras, 445, 1745

\bibitem[{{Yoshiura} {et~al}\mbox{.}(2016){Yoshiura}, {Hasegawa}, {Ichiki},
  {Tashiro}, {Shimabukuro}, \& {Takahashi}}]{2016arXiv160204407Y}
{Yoshiura} S., {Hasegawa} K., {Ichiki} K., {Tashiro} H., {Shimabukuro} H.,
  {Takahashi} K., 2016, ArXiv e-prints

\end{thebibliography}

\appendix

\section{Model Uncertainties}
\label{SEC:model_uncertainties}

In this section we discuss caveats for our models of the Ly$\alpha$ forest opacity fluctuations considered in \S \ref{SEC:opacityflucs}.  Below we argue that our models do not account for a number of effects that would, in combination, likely lead to a net decrease in the amplitude of $\taueff$ fluctuations: 

\begin{enumerate}

\item
We do not include the effects of inhomogeneous heating from patchy \HI\ reionization.  These temperature inhomogeneities would suppress fluctuations in $\taueff$ because the overdense regions that were reionized earlier would tend to be cooler than average by $z\sim5.5$ (i.e. after reionization), with enhanced equilibrium \HI\ densities from the $\propto T^{-0.7}$ dependence of the recombination coefficient.  However, the ionizing background in those regions would tend to be stronger owing to the bias of the sources.  The competition between these two effects would smooth out fluctuations in $\taueff$ (recall that $\tau_{\mathrm{Ly}\alpha} \propto T^{-0.7} \Delta^2/\GammaHI$).  If \HI\ reionization ends at $z\approx6$, this competition could have a significant effect on the width of the $\taueff$ distribution \citep{2015ApJ...813L..38D}.  On the other hand, we also do not account for inhomogeneous heating during \HeII\ reionization, which would likely work in the direction of \emph{enhancing} the $\taueff$ fluctuations, since the regions of strong ionizing background surrounding AGN would also be the most recently photoheated. 

\item
We do not take into account the finite lifetimes of AGN.  With a duty cycle of unity, our models effectively assume an infinite AGN lifetime.  We argue that modeling the finite lifetimes could act to suppress $\taueff$ fluctuations in at least two ways.  First, for a fixed AGN abundance (which is set by the assumed luminosity function and magnitude limit), a smaller duty cycle would require that AGN are hosted in halos that are less-massive and therefore less biased than in our models.  This would lead to the AGN being less clustered overall, which would yield somewhat smaller $\taueff$ fluctuations.  Secondly, consider a region with an overdensity of bright AGN in our fiducial model (see e.g. the lower left corners of the panels in Fig. \ref{FIG:visB}).  These regions produce large segments of the Ly$\alpha$ forest with low optical depth.  Suppose now that only a fraction of these AGN are active at a given redshift.  In this case, the Ly$\alpha$ opacity in this large-scale region would be less coherent -- an effect that would make the $\taueff$ distribution narrower.  On the other hand, we note that the proximity zone remnants left by AGN that recently became dormant might mitigate these effects. 

\item
We do not account for the anisotropy of emissions from AGN.  Our argument for why the beaming of the radiation from AGN would lead to lower $\taueff$ fluctuations is similar to our discussion at the end of the last paragraph.  Consider again a region with an overdensity of AGN.  In our current model, these sources are treated as isotropic emitters.  If instead the ionizing radiation was beamed in random directions, the coherence scale of the opacity in that region would be smaller, leading to smaller fluctuations in $\taueff$.

\item
Lastly, we have neglected the effect of recombination radiation, which would act to homogenize the ionizing background,  leading to smaller fluctuations in $\taueff$.  However, we note that this effect is likely minor for our galaxies$+$AGN models, where the contribution from galaxies is already relatively homogeneous owing to the large mean free path.  Recombination radiation may play a more significant role in galaxies-only models where the mean free path is short.   

\end{enumerate}

We expect that the above effects in combination would lead to a net decrease in the amplitude of $\taueff$ fluctuations in our models.  Thus accounting for the observed $\taueff$ fluctuations would require a larger contribution from AGN.  According to our findings in \S \ref{SEC:thermalhistory}, this would only create more tension between these model and current constraints on \HeII\ reionization and on the thermal history of the IGM. 

\section{AGN luminosity functions used in this work}
\label{SEC:simLFs}

In this section we present details on the AGN populations for the models presented in \S \ref{SEC:opacityflucs}.  The luminosity function of AGN is typically parameterized with a double power law,

\begin{equation}
\Phi(M) = \frac{\Phi^*}{10^{0.4 (M_{\rm break} - M) (\beta - 1)} + 10^{0.4 (M_{\rm break} - M) (\gamma - 1)}},
\end{equation}
where $M$ is magnitude, $\beta$ and $\gamma$ respectively correspond to the logarithmic slopes of the faint and bright ends of the luminosity function, and $M_{\rm break}$ marks the turnover.  For the luminosity function measurement of G2015, $\log \Phi^{*} = -5.8$, $\beta = 1.66$, $\gamma = 3.35$, and $M_{\rm break} = -23.4$ (AB, 1450 \AA) in their highest redshift bin of $z=5-6.5$.  

We use these values as a baseline for our AGN-driven models of the $z>5$ ionizing background, rescaling $\Phi^{*}$ to explore scenarios with varying levels of contribution from AGN.  Since the redshift evolution of the AGN abundance reported by G2015 is rather modest across $z\sim4-6$, we assume no evolution in the luminosity function across $z=5-5.7$, i.e. the redshifts relevant for the opacity fluctuation measurements of \citet{2015MNRAS.447.3402B}. For our $25\%$, $50\%$, and $90\%$ AGN models (in which AGN emissions contribute $25\%$, $50\%$, and $90\%$ of the spatially averaged \HI\ photoionization rate, $\langle \GammaHI \rangle$), we adopt the normalizations of $0.5\Phi^{*}$, $\Phi^{*}$, and $1.8\Phi^{*}$, respectively.  (Note that the $50\%$ model is exactly the luminosity function of G2015.)  We randomly draw luminosities from these rescaled luminosity functions to populate with AGN the most massive dark matter halos in our composite simulation box.  For our volume of $(400h^{-1}~\Mpc)^3$, this yields $3,506$, $7,012$ and $12,622$ AGN for the $25\%$, $50\%$, and $90\%$ models, respectively.  During our computations of the fluctuating ionizing background, the escape fraction of galaxies is tuned to yield the remaining percentage of $\langle \GammaHI \rangle$.   The normalizations given above were chosen such that in each model the total $\langle \GammaHI \rangle$, including the contribution from galaxies, is approximately the value that is required to match the observed transmission in the forest.  During the final steps of constructing the $\taueff$ distribution, $\langle \GammaHI \rangle$ is then rescaled by a small amount ($\lesssim 10\%$) to match the observed transmission. 

\begin{figure}
\begin{center}
\resizebox{8.5cm}{!}{\includegraphics{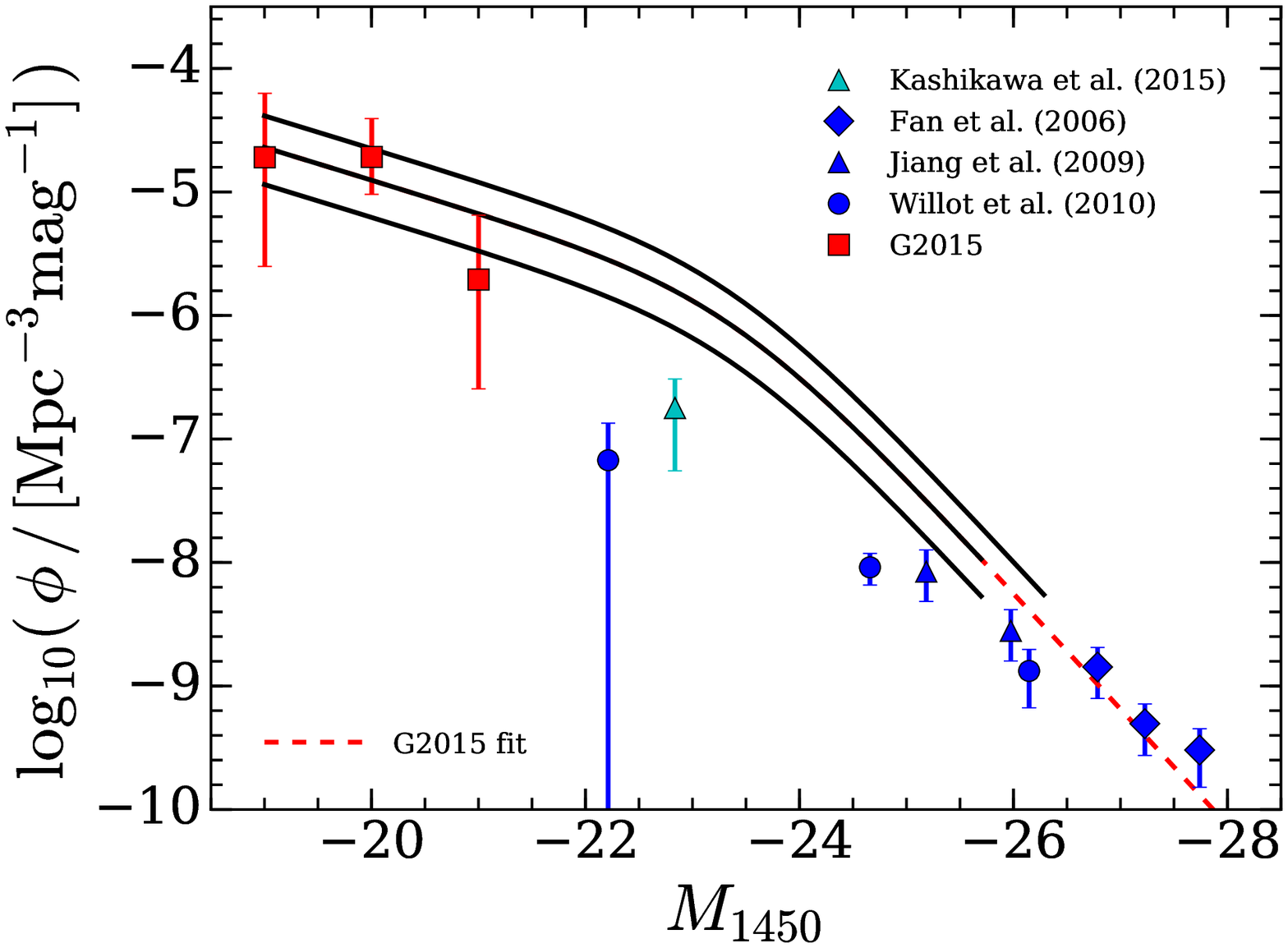}}
\end{center}
\caption{AGN luminosity functions used in our models of the AGN-driven ionizing background.  From bottom to top, the black/solid curves correspond to models in which AGN emissions contribute $25\%$, $50\%$, and $90\%$ of the spatially averaged \HI\ photoionization rate, $\langle \GammaHI \rangle$. The left(right) end of the curves terminate at the lowest(highest) luminosity sampled in our simulations.  For reference, the red/dashed curve shows the luminosity function measurement of G2015, while the data points show a compilation of recent observational measurements.    }  
\label{FIG:simLF}
\end{figure}

From bottom to top, the black/solid curves in Fig. \ref{FIG:simLF} show the luminosity functions of AGN for the $25\%$, $50\%$, $90\%$ models.  The left(right) end of the curves terminate at the lowest(highest) luminosity sampled in our simulations.  For comparison the red/dashed curve shows the luminosity function measurement of G2015, while the data points show the compilation of recent observational measurements presented in Fig. \ref{FIG:lumfunc} of the main text.

\section{Ionizing emissivity models}
\label{SEC:emissivitymodels}

In \S \ref{SEC:thermalhistory}, we quantified the effects of a large $z>5$ AGN population on the thermal history of the IGM. In this section we provide the formulae used to model the \HI\ ionizing emissivity from the AGN population.  For the ``standard" scenario based on previous optical AGN surveys (in which the AGN abundance drops so rapidly that galaxies dominate the $z>4$ \HI\ ionizing background), we use a modified version of the widely used \citet{2012ApJ...746..125H} model for the comoving volume emissivity at $\lambda = 912\AA$, 

\begin{equation}
\epsilon^{\rm AGN}_{912}(z) = 12\times 10^{24} \left( \mathrm{\frac{erg/s}{Mpc^{3}~Hz}} \right)  \frac{(1+z)^4 \exp\left(-0.2z\right)}{ \exp\left(1.9z\right) + 27.3 }.
\label{EQ:AGNemStandard}
\end{equation}
We find equation (\ref{EQ:AGNemStandard}) to be a better match to the AGN emissivity for the most recent measurements at both low and high redshifts. For the other scenarios in which AGN contribute significantly to the high-$z$ ionizing background, we adopt a modified form of the recently published \citet{2015ApJ...813L...8M} model,

\begin{equation}
\log_{10} \epsilon^{\rm AGN}_{912}(z) = C_1 \exp\left( -C_2 z \right) - C_3 \exp(-C_4 z),
\label{EQ:modMH2015}
\end{equation}
where $(C_1, C_2, C_3, C_4) = (25.75, 0.0088, 2, 0.8)$ and $(25.9,0.013,2.2,0.95)$ respectively for the red/solid and green/long-dashed curves in Figs. \ref{FIG:AGNemissivityB} and \ref{FIG:temphistories}.

\end{document}